\documentclass[aps,prx,reprint]{revtex4-2}
\usepackage[T1]{fontenc}
\usepackage{xcolor}
\usepackage{babel}
\usepackage{mathtools}
\usepackage{amsmath}
\usepackage{amssymb}
\usepackage{graphicx}
\usepackage[unicode=true,pdfusetitle,
 bookmarks=true,bookmarksnumbered=false,bookmarksopen=false,
 breaklinks=true,pdfborder={0 0 1},backref=false,colorlinks=true]
 {hyperref}
\hypersetup{
 pdfborderstyle=,citecolor=blue}

\makeatletter
\usepackage{babel}
\usepackage{array}
\usepackage{multirow}
\usepackage{times}
\usepackage{longtable} 

\makeatother

\makeatletter
\let\oldaddcontentsline\addcontentsline
\newcommand{\stoptocentries}{\renewcommand{\addcontentsline}[3]{}}
\newcommand{\starttocentries}{\let\addcontentsline\oldaddcontentsline}
\makeatother

\begin{document}

\title{Nonequilibrium Theory for Molecular Machine Design}
\author{Ying-Jen Yang}
\email{ying-jen.yang@stonybrook.edu}
\affiliation{Laufer Center of Physical and Quantitative Biology, Stony Brook University}
\author{Ken A. Dill}
\affiliation{Laufer Center of Physical and Quantitative Biology, Stony Brook University}
\affiliation{Department of Physics and Astronomy, Stony Brook University}
\affiliation{Department of Chemistry, Stony Brook University}
\date{\today}

\begin{abstract}
Modeling the dynamical flows on networks of biomolecular machines often entails computing node populations and edge fluxes with Master Equations and correlating machine performance with entropy production.  But this alone is not sufficient for design, optimization and evolution because it doesn't treat cost-benefit tradeoffs, or small-system misflows (backsteps, futile cycles, ineffective actions), or differential properties for flow design. Here we develop CFT Design, based on the recently developed Caliber Force Theory (CFT). We apply it to: designing faster molecular motors through ``traffic control''; optimizing speed, energy, and accuracy in kinetic proofreaders; and designing better enzyme inhibitors. CFT Design provides a general framework for optimizing nonequilibrium flow networks.
\end{abstract}

\maketitle
\stoptocentries

\section*{Network Flows need design principles.}
We are interested in network flows---fluids, particles, energies; trafficking of vehicles or goods; ecological movements of organisms and species; and especially the cyclic processes of biomolecular machines. These dynamics are commonly modeled by stochastic dynamical models where a traveling agent ``rolls a die'' to determine when and where to transition. For simple models of biomolecular machines, this die roll comes from the thermal noise of climbing energy barriers, described by Arrhenius-type exponential waiting times. Today's modeling computes probability distributions and edge flows using Master Equations or Markov State Models, often focusing on correlating machine performance with entropy production~\cite{qian_phosphorylation_2007, horowitz_thermodynamics_2014, parrondo_thermodynamics_2015, seifert_stochastic_2019, peliti_stochastic_2021, cao_stochastic_2025, leighton_flow_2025, floyd_limits_2025, tu_nonequilibrium_2026}.\\

What is missing for optimizing and designing flow networks?  First, we need a way to treat the costs and benefits of flows.  Whereas equilibrium (EQ) thermodynamics often balances the tradeoffs of heat and work, for example in Carnot cycles, network flow streams can have a much broader range of desirability properties.  Product supply chains have dollar costs and benefits; biochemical pathways produce valued biomolecules and waste products; biomolecular proofreading trades off speed, accuracy and energy costs; etc.  And, networks have time costs, which depend not only on how fast an agent takes a local transition, but also on slowdowns due to wrong turns, futile cycles, or dead ends. Network flow design theory must handle these more global \textit{topological misflows}~\cite{qian_metabolic_2006,baiesi_life_2018,maes_frenesy_2020}.\\

Second, non-equilibrium (NEQ) physics has not had the same level of powerful mathematical relationships that EQ has had, 
for example, establishing analytical relationships that connect flow variables to forces and constraints across different networks.
This contrasts with EQ thermodynamics mathematics which relates complete variable sets, such as $(U, V, N)$, to their conjugate forces $(T, p, \mu)$ and to whatever are the constraints that are imposed upon the system in the problem at hand. Here, we develop principles for the design, optimization, and evolution of flows under constraints. Based on the recent Caliber Force Theory (CFT) \cite{yang_principled_2026,yang_fluctuation-response_2026}, we establish the observable-force conjugacy for NEQ dynamics to formalize a framework of flow design. We apply the framework to three biomolecular machines with distinct types of misflows: correcting backstepping in the F$_1$-ATPase motor \cite{gerritsma_chemomechanical_2010}; optimizing branching across incorporation or correction cycles in kinetic proofreaders \cite{ninio_kinetic_1975,hopfield_kinetic_1974,banerjee_elucidating_2017,mallory_trade-offs_2019}; and strengthening dead-ends or futile cycles in enzyme inhibition \cite{cleland_kinetics_1963,copeland_evaluation_2005,cornish-bowden_fundamentals_2012}.

\section*{Caliber Forces give Flow Design Rules.}
\begin{figure}
    \centering
    \includegraphics[width=0.95\linewidth]{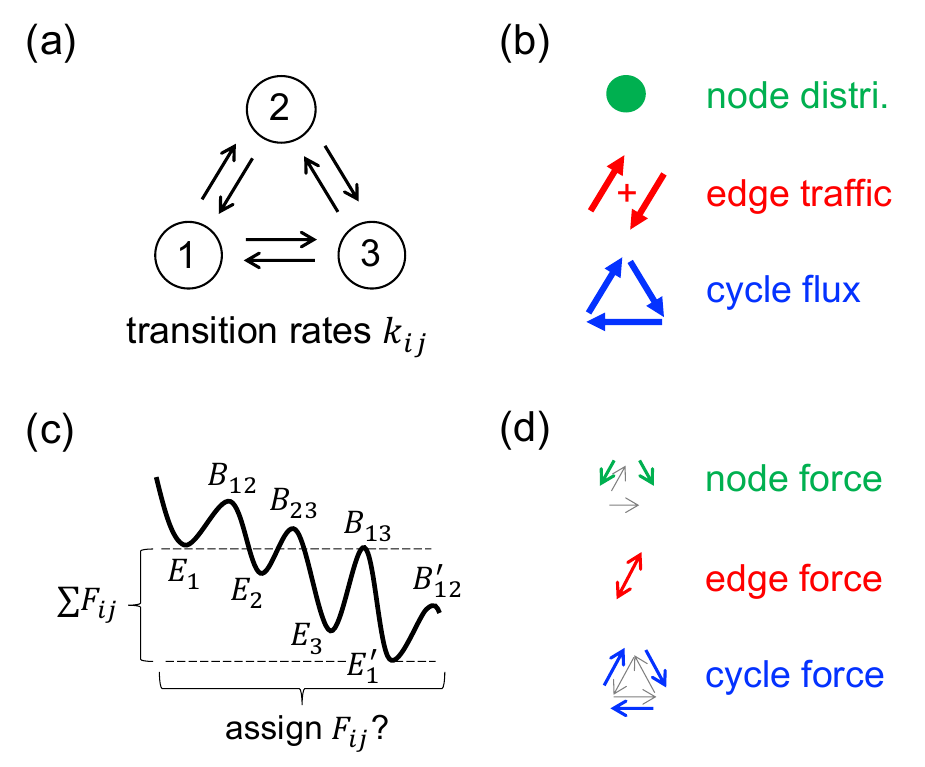}
    \caption{ \textbf{A simple 3-state machine process, with CFT properties defined.} \textbf{(a)} A Markov network . \textbf{(b)} The observables: node probability distributions, symmetric edge traffics, and directed cycle fluxes. \textbf{(c)} Commonly modeled using an Arrhenius energy landscape ($E_n, B_{ij}$) and a cyclic drive ($\sum_{(ij)\in c}F_{ij}$) that tilts the landscape. The primed variables are the unprimed shifted by $\sum F_{ij}$. There is a gauge ambiguity in distributing the drive onto each edge---many combinations of $F_{ij}$ and $E_n$ lead to the same dynamics. \textbf{(d)} CFT resolves this ambiguity by deriving conjugate forces as affinities to node dwelling, edge exchange, and cycle completion.}
    \label{fig: EBF versus forces}
\end{figure}

To begin, we first summarize the relevant parts of Caliber Force Theory (CFT), a principled general treatment of non-equilibria. Consider a molecular machine modeled as a Markov jump process on a fixed network state space (e.g., Fig. \ref{fig: EBF versus forces}a). As illustrated in Fig. \ref{fig: EBF versus forces}b, the machine's long-term functional performance is characterized by its steady-state statistics, governed by three types of observables \cite{yang_principled_2026}: the node probabilities $\pi_i$, the symmetric edge traffics \cite{maes_frenesy_2020} $\tau_{ij} = \pi_i k_{ij} + \pi_j k_{ji}$ where $k_{ij}$ is the transition rate from state $i$ to $j$, and the antisymmetric edge net fluxes $J_{ij}=\pi_i k_{ij} - \pi_j k_{ji}$ spanned by the set of fundamental cycle fluxes $J_c$ (Kirchhoff's current law) \cite{schnakenberg_network_1976,hill_free_1989}. The goal is to control and optimize these observables by exploring the parameter space of transition rates ($k_{ij}$).\\

In some situations, molecular machines are driven by differences across two or more equilibrium reservoirs and the flows are constrained by the Local Detailed Balance (LDB) condition \cite{seifert_stochastic_2018,maes_local_2021,falasco_local_2021}.  This fixes the cycle affinity of any fundamental cycle $c$: 
\begin{equation}
\sum_{(ij)\in c}\ln\frac{k_{ij}}{k_{ji}} = \frac{\Delta\mu - w}{k_\text{B} T},
\end{equation}
where $\Delta \mu$ is the free energy consumption after completing the cycle and $w$ is the work done by the cycle. To separate the controllable (tunable) degrees of freedom from these thermodynamic constraints, a common approach is to assume Arrhenius behavior of the transition rates  \cite{owen_universal_2020}:
\begin{equation} \label{eq: Arrhenius parameterization}
    k_{ij} = \mathcal{N}\exp[-(B_{ij}-E_i)]\exp(F_{ij}/2),
\end{equation}
where $\mathcal{N}$ is a global timescale prefactor, $E_i$ are node energies ($k_{\text{B}}T$ unit), $B_{ij}$ are kinetic edge barriers, and $F_{ij}$ are local edge affinities distributing the fixed cycle affinity. While this parameterization (Fig. \ref{fig: EBF versus forces}c) has been useful in discovering various response relations \cite{owen_universal_2020,fernandes_martins_topologically_2023,aslyamov_nonequilibrium_2024}, these parameters $E$, $B$, and $F$ lack a conjugate structure with respect to the dynamical observables $\bar{\textbf{x}}=(\boldsymbol{\pi}, \boldsymbol{\tau}, \boldsymbol{J})$ (see SI Sec. I.A \cite{SI}). To get general design principles, resembling those that underlie the power of EQ thermodynamics, we must first identify the true conjugate forces of NEQ dynamics. \\

CFT derives the conjugate forces in terms of the maximization of the path entropy (Maximum Caliber) of the Markov dynamics \cite{jaynes_minimum_1980,presse_principles_2013,yang_principled_2026}. The forces conjugate to the node distribution $\pi_n$, the edge traffic $\tau_{ij}$, and the cycle fluxes $J_c$ represent the affinities for node dwelling, edge exchange, and cycle completion, respectively (Fig. \ref{fig: EBF versus forces}d):
\begin{align}
\mathfrak{F}_{\text{node},n} & = \sum_{i(\neq m)}(k_{mi}-1)-\sum_{j(\neq n)}(k_{nj}-1),\nonumber \\
\mathfrak{F}_{\text{edge},ij} & = \frac{1}{2}\ln{k_{ij}k_{ji}},\nonumber \\
\mathfrak{F}_{\text{cycle},c} & = \frac{1}{2}\sum_{(ij)\in c}\ln\frac{k_{ij}}{k_{ji}}. \label{eqs: forces}
\end{align}
CFT handles the LDB constraint naturally as $\mathfrak{F}_{\text{cycle},c}$ is half of the cycle affinity, and these forces enable machine design.\\\\

\textbf{Conjugate coordinates solve constrained optimization.} 
Machine design first requires determining the optimal performance under physical constraints (e.g., maximizing speed under fixed free energy driving). CFT solves this using its conjugate coordinate systems and Legendre transformations \cite{yang_principled_2026}. By shifting from the observable coordinate $(\boldsymbol{\pi},\boldsymbol{\tau},\boldsymbol{J})$ to the observable-force coordinate $(\boldsymbol{\pi},\boldsymbol{\tau},\boldsymbol{\mathfrak{F}}_{\text{cycle}})$ (see SI Sec. I.A \cite{SI}), we can simply hold the environmental cycle forces $\boldsymbol{\mathfrak{F}}_{\text{cycle}}$ constant while tuning the remaining parameters to meet functional targets and solve the underlying optimal rates $k_{ij}$. This has yielded the \textit{Equal Traffic Principle} \cite{yang_principled_2026}---the rule that maximizing cycle flux in a unicircular process under a fixed total traffic budget requires maintaining equal traffic across all steps. Here, we apply this principle to diagnose the efficiency of the F$_1$ molecular motor, extend it to design stronger enzyme inhibition, and show how to tune the speed, accuracy, and cost of kinetic proofreading networks.\\

\textbf{Caliber forces yield flow response rules.}
The second task in machine design is to predict how tuning physical parameters affects the routing of network flows. CFT establishes far-from-EQ generalizations to Maxwell-Onsager symmetries and fluctuation-response equalities, mapping how dynamic observables respond to force perturbations \cite{yang_principled_2026,yang_fluctuation-response_2026}. While these universal force principles do not require an Arrhenius assumption, biomolecular machines are physically tuned by evolution through their specific energy landscapes. By projecting these specific evolutionary knobs ($E_n, B_{ij}$) onto the general CFT forces (detailed in \cite{yang_fluctuation-response_2026} and SI Sec. I.B \cite{SI}), we obtain two explicit design rules for flow routing via Arrhenius parameters.
\begin{enumerate}
    \item \textbf{Node Energies affect the scaling of fluxes:} CFT shows that node energy perturbations isotropically scale the average value of any flux variable $\bar{\varphi} \in (\tau_{ij}, J_c)$ (or their linear combinations):
    \begin{equation} \label{eq: d flux/ dE}
        \frac{\partial \bar{\varphi}}{\partial E_n} = \pi_n~ \bar{\varphi}.
    \end{equation}
    The state (node) with highest $\pi_n$ is the most effective scaler, but pure $E_n$ perturbations cannot alter flux ratios. Thus, node energies alone cannot change the accuracy of kinetic proofreading networks \cite{mallory_kinetic_2020}. This also yields a new scaling relation governing how node energy modulates the entropy production rate (EPR) of the network:
    ${\partial_{E_n} \text{EPR}} \equiv \partial_{E_n}[\sum_c J_c ~(2\mathfrak{F}_{\text{cycle},c})] = \pi_n ~\text{EPR}$
    where we have used $2\mathfrak{F}_{\text{cycle},c} = \sum_{(ij)\in c} \ln (k_{ij}/k_{ji}) = \sum_{(ij)\in c} \text{arctanh} (J_{ij}/\tau_{ij})$, confirming that its derivative with respect to $E_n$ vanishes.
    \item \textbf{Kinetic Barriers affect the routing of fluxes:} Conversely, barrier perturbations $B_{ij}$ route fluxes, mediated by the edge's net flux $J_{ij}$. For any distribution or flux observable $\bar{x}_\alpha\in(\pi_n,\tau_{ab},J_c)$ (except the local traffic $\tau_{ij}$, which has an additional term in the equation \cite{SI}): \begin{align} \label{eq: dx/dB}
    \frac{\partial \bar{x}_\alpha}{\partial B_{ij}} 
    &= -J_{ij}~ A^{-1}_{\alpha,(ij)} 
    \end{align} where $\mathbf{A}^{-1}$
      is the inverse of the force-rate Jacobian matrix $\mathbf{A}_{(ij),\alpha}\equiv k_{ij} \partial_{k_{ij}}\mathfrak{F}_\alpha$, which can be easily evaluated based on Eqs. \eqref{eqs: forces}
    \cite{yang_fluctuation-response_2026,SI}. If an edge has zero net flux ($J_{ij}=0$)---such as those leading to dead-end pathways---tuning its barrier alters only its own traffic $\tau_{ij}$ while leaving all other network observables invariant. We utilize this topological routing rule to  establish the design limits of enzyme inhibitors.\\
\end{enumerate}

These relationships constitute a comprehensive framework for molecular-machine \textit{inverse design}  (solving constrained optimal rates via conjugate coordinates) and \textit{forward design} (routing flows via Arrhenius response rules). Here are three examples.

\section*{1) Speeding up motors through traffic control}

\begin{figure}
    \centering
    \includegraphics[width=0.9\linewidth]{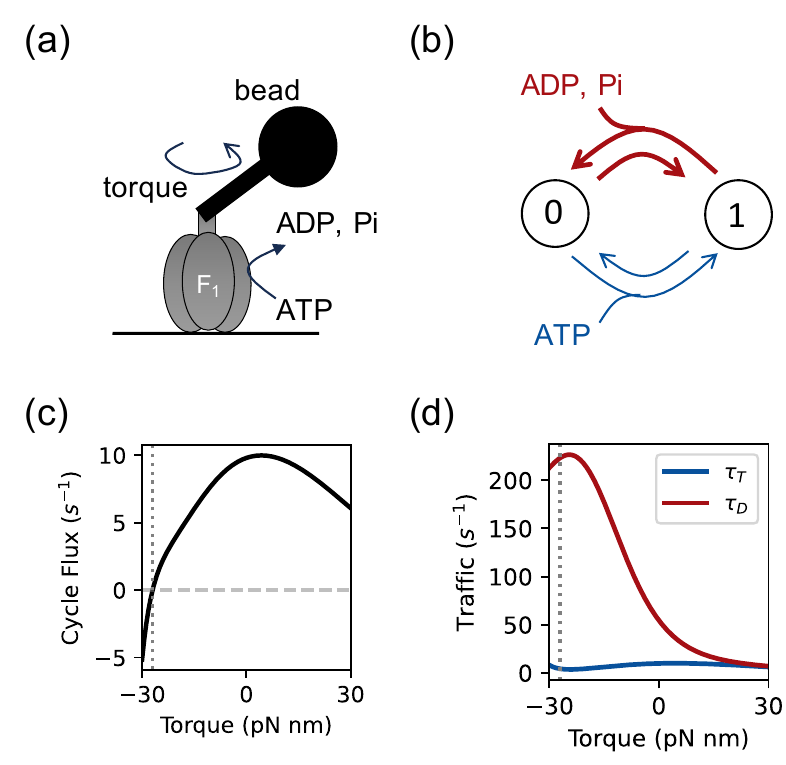}
    \caption{\textbf{The F$_1$-ATPase rotary motor properties.} (a) In the Gerritsma and Gaspard assay, the motor is fixed to the surface and the bead rotates \cite{gerritsma_chemomechanical_2010}. (b) The two-state Markov network: ATP binding (blue) and catalysis (red) transitions. The F$_1$ motor rotates $120^\circ$ every time a 0-1-0 (de-) hydrolysis cycle is completed. (c) The steady-state cycle flux, and (d) the symmetric traffics on the two edges versus the applied external torque. The traffic imbalance ($\tau_D \gg \tau_T$) indicates the motor's processive inefficiency \textit{in vitro}.}
    \label{fig: F1}
\end{figure}

The F$_o$-F$_1$ ATP synthase is biology's rotary transducer: the F$_o$ motor converts the downhill flow of protons into mechanical torque, which drives the F$_1$ motor to perform the energetically uphill synthesis of ATP. While earlier models evaluate its \textit{thermodynamic efficiency} (work output versus free energy consumed) \cite{wagoner_mechanisms_2019,brown_theory_2020}, we show here that its speed and utility are also constrained by its \textit{processive efficiency}, defined as the ratio of net transport to total transitions taken ($J/\tau$). Here, we show that the F$_1$ motor is limited by futile backstepping, a traffic imbalance that can only be corrected by routing flows via kinetic barriers.\\

To model the \textit{in vitro} F$_1$ motor assay (Fig.~\ref{fig: F1}a), we adopt the 2-state, 2-edge network from Gerritsma and Gaspard \cite{gerritsma_chemomechanical_2010} (Fig.~\ref{fig: F1}b). Under the low-ATP conditions in the experiments ($\approx 0.5\ \mu$M), the slow ATP-binding rate justifies lumping the subsequent transitions. The motor transitions between an empty (node 0) and a nucleotide-occupied state (node 1) via the T transition (ATP binding/unbinding, $\approx 90^\circ$ rotation) and the D transition (catalysis and product release, $\approx 30^\circ$ rotation). The model fits well to the experimental data.\\

The network dynamics are governed by four load-dependent transition rates, $k_{\text{T}\pm}(\Gamma)$ and $k_{\text{D}\pm}(\Gamma)$, where $\Gamma$ is the applied torque (see SI Sec. II.A \cite{SI}). We confine our analysis to $\Gamma \in [-30, 30]$ pN nm to preclude mechanical slipping and maintain the tight-coupling condition. These phenomenological rates satisfy the LDB cycle force constraint:
\begin{equation}
    \ln \frac{k_{\text{T}+}(\Gamma)k_{\text{D}+}(\Gamma)}{k_{\text{T}-}(\Gamma)k_{\text{D}-}(\Gamma)} = 2\mathfrak{F}_{\text{cycle}} = \frac{\Delta \mu + \Gamma \Delta\theta}{k_{\text{B}} T},
\end{equation}
where $\Delta\theta = 2\pi/3$ is the rotation angle, and $\Delta \mu$ is the chemical free energy of ATP hydrolysis. Of interest are the steady-state distribution of the two states ($\pi_0, \pi_1$), the edge traffics ($\tau_{\text{T}}, \tau_{\text{D}}$), and the cycle flux ($J_c$) representing 1/3 of the motor's physical rotation speed.\\

CFT's observable-force coordinate $(\pi_1,\tau_\text{T},\tau_{\text{D}},\mathfrak{F}_{\text{cycle}})$ diagnoses the motor's processive inefficiencies by decoupling the constrained free energy drive ($\mathfrak{F}_{\text{cycle}}$) from the tunable kinetic variables. Under a fixed total kinetic activity budget ($\tau_{\text{tot}} = \tau_\text{T} + \tau_\text{D}$), maximizing the cycle flux $J_c$ requires the traffic to be balanced across all steps ($\tau_{\text{T}} = \tau_{\text{D}}$) \cite{yang_principled_2026}. This \textit{Equal Traffic Principle} (ETP) bounds the motor's processive efficiency ($J_c/\tau_{\text{tot}}$) (see SI Sec. II.B \cite{SI}):
\begin{equation}
    \frac{J_c}{\tau_{\text{tot}}} \le \frac{1}{2} \tanh\left( \frac{\mathfrak{F}_{\text{cycle}}}{2} \right).
\end{equation}
As shown in Fig.~\ref{fig: F1}d, the \textit{in vitro} rotor operates far below this optimum. The traffic imbalance ($\tau_{\text{D}} \gg \tau_{\text{T}}$) indicates that kinetic activity is wasted on futile backstepping along the D transition.\\

To correct this imbalance, we compute the energetic response relations (Eqs.~\ref{eq: d flux/ dE}--\ref{eq: dx/dB}) with the analytical Jacobian inverse (see SI Sec. II.C \cite{SI}):
\begin{subequations} \label{eqs: dlnJ/dEB}
   \begin{align}
   \frac{\partial \ln J_c}{\partial E_n} &= \frac{\partial \ln \tau_\text{T}}{\partial E_n}=\frac{\partial \ln \tau_\text{D}}{\partial E_n} = \pi_n;\\
   \frac{\partial \ln J_c}{\partial B_{\text{T}}} &= -\frac{\Sigma_{\text{D}}}{\Sigma}, \quad
   \frac{\partial \ln J_c}{\partial B_{\text{D}}} = -\frac{\Sigma_{\text{T}}}{\Sigma},
   \end{align}
\end{subequations}
where $\Sigma_{\text{X}} = k_{\text{X}+} + k_{\text{X}-}$ and $\Sigma$ is the sum of all rates. These relations dictate that improving efficiency requires using the kinetic barriers ($B_{\text{T}}, B_{\text{D}}$) as topological routers to balance the traffic. Conversely, perturbing the node energy $E_n$ only scales the fluxes by $\pi_n$, shifting state occupancies without altering the traffic ratio ($\tau_{\text{T}}/\tau_{\text{D}}$) or the processive efficiency---a structural degeneracy inherently mapped by the $(\pi_1,\tau_\text{T},\tau_{\text{D}},\mathfrak{F}_{\text{cycle}})$ coordinate (Appendix II.B of SI \cite{SI}). Notably, these barrier sensitivities sum to $-1$, manifesting a cycle response symmetry \cite{yang_fluctuation-response_2026} analogous to the Flux Control Summation Theorem in Metabolic Control Analysis \cite{kacser_control_1995,heinrich_linear_1974}.\\

In summary, the \textit{in vitro} F$_1$-ATPase operates far from its optimal processive efficiency, due to considerable backstepping. While node energies scale the motor activity within a degenerate space of state occupancies, the kinetic barriers affect the routing and can correct this traffic imbalance.

\section*{2) Decoupling speed, accuracy, and cost in Kinetic Proofreading}

Hopfield \cite{hopfield_kinetic_1974} and Ninio \cite{ninio_kinetic_1975} were the first to explain how biomolecular processes can harness non-equilibrium driving to achieve certain biochemical process accuracies beyond the limits of thermodynamic equilibrium. For example, in \textit{kinetic proofreading} (KP), a DNA polymerase enzyme copies DNA sequences onto new DNA with extremely small error rates, of one in $10^9$ monomers. Traditionally, KP is viewed as a trade-off among speed, accuracy, and energy cost \cite{murugan_speed_2012,sartori_thermodynamics_2015}. However, our design framework reveals that these canonical trade-offs are not absolute; they only manifest at the boundaries of the system's kinetic capacity. Within the operational interior, these performance metrics can be effectively decoupled. We show here how kinetic barriers act as topological routers to circumvent these compromises, and remarkably, how real proofreading networks have evolved quite close to this optimal saturation point.\\

\begin{figure}
    \centering
    \includegraphics[width=\linewidth]{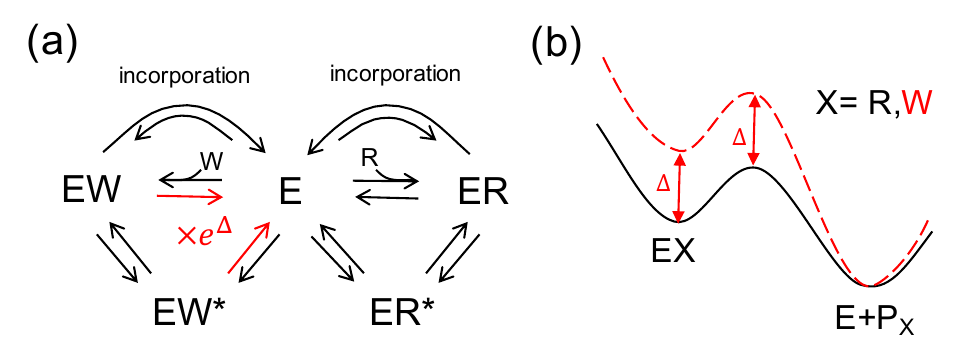}
    \caption{\textbf{The classical model of kinetic proofreading assumes barrier asymmetry.} (a) Traditional models assume that discriminating Right (R) from Wrong (W) relies solely on accelerating the unbinding of the W substrate, while keeping forward incorporation rates identical. (b) Because the Wrong bound state (red dashed) is destabilized by a node energy $\Delta$, maintaining identical incorporation rates mathematically requires elevating the Wrong kinetic barrier by exactly the same $\Delta$.}
    \label{fig: KP}
\end{figure}

The topological necessity of barrier routing is embedded in foundational KP models \cite{ninio_kinetic_1975,hopfield_kinetic_1974}. Classic models  attributed error correction primarily to differing unbinding rates for the Right (cognate) and Wrong (non-cognate) substrates, while assuming identical forward incorporation rates (Fig.~\ref{fig: KP}a). However, applying CFT response rules reveals a hidden topological constraint. As established earlier, node energies ($E_n$) only scale fluxes ($\partial \ln J_c / \partial E_n = \pi_n$). Thus, node energy differences alone cannot alter the error rate ($\varepsilon = \log_{10}(J_{\text{W}}/J_{\text{R}})$), as independently reported by others \cite{mallory_kinetic_2020}. To maintain identical incorporation rates while altering unbinding rates, the classic model implicitly requires asymmetric incorporation barriers (Fig.~\ref{fig: KP}b). Barrier routing, not just binding affinity, is the engine of proofreading.\\

To move beyond idealized symmetric models, we apply our framework to the asymmetric networks of real KP systems mapped by recent single-molecule experiments: the \textit{E. coli} ribosome and T7 DNA polymerase \cite{banerjee_elucidating_2017,mallory_trade-offs_2019}.\\

\begin{figure}
    \centering
    \includegraphics[width=\linewidth]{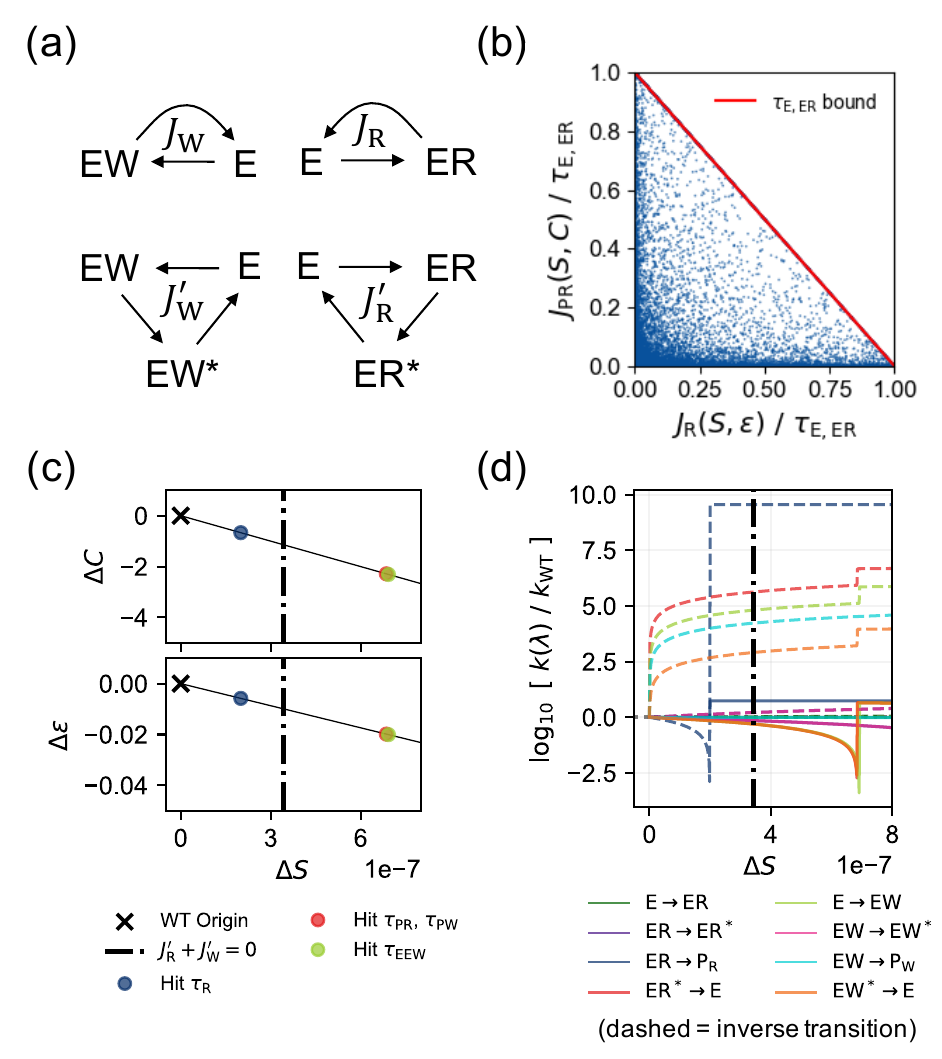}
    \caption{\textbf{Inverse design of the T7 DNA polymerase.}
\textbf{(a)} The four fundamental cycles defining the incorporation ($J_{\text{R}}, J_{\text{W}}$) and proofreading ($J'_{\text{R}}, J'_{\text{W}}$) fluxes.
\textbf{(b)} Validating the trade-off bound in Eq. \eqref{eq: KP_bounds} where $J_{\text{R}}$ and $J'_{\text{R}}$ are functions of the speed ($S$), accuracy ($\varepsilon$), cost ($C$). See main text for expressions. Dots are generated via $10^5$ log-uniform traffic ($\tau$) values $\in (10^{-3},10^{5})$. Bound saturates when $J_\text{R}+J'_{\text{R}}=\tau_{\text{E,ER}}\Leftrightarrow p_{\text{ER,E}}=0$.
\textbf{(c)} An inversely-designed ``evolutionary'' process increasing speed $S$ while reducing error $\varepsilon$ and cost $C$ from the wild-type origin (black cross). $\Delta Z=Z-Z_{\text{WT}}$ represents the excess to wild-type values ($Z=C,\varepsilon,S$). Colored dots indicate where specific traffic limits ($\sum J \le \tau$) saturate and are enlarged tenfold to allow further optimization. The vertical dash-dotted line marks the transition into the anti-proofreading regime ($J'_{\text{W}}+J'_{\text{R}}\le 0$).
\textbf{(d)} Solving the driving protocol of transition rates realizing the linear inverse design path in (c). It requires speeding up various reversed rates (colored dashed line). Jumps reflects the traffic limits relaxation algorithm. }
\label{fig: KP inverse design}
\end{figure}

\textbf{Performance trade-offs emerge only at the parameter-space boundaries.}
We use CFT's coordinate systems to show that speed-accuracy-cost trade-offs are boundary constraints. To demonstrate this in the T7 DNA polymerase, we formalize the network's performance using four fundamental cycle fluxes (Fig.~\ref{fig: KP inverse design}a): correct and incorrect incorporations ($J_{\text{R}}, J_{\text{W}}$) and their corresponding proofreading discards ($J'_{\text{R}}, J'_{\text{W}}$). The performance metrics are algebraic combinations of these fluxes: total incorporation speed $S = J_{\text{R}} + J_{\text{W}}$, error $\varepsilon = \log_{10} (J_{\text{W}} / J_{\text{R}})$, and power cost $C = S\Delta\mu_\text{p} + (J'_{\text{R}} + J'_{\text{W}})\Delta\mu_{\text{NTP}}$, where $\Delta\mu_\text{p}$ and $\Delta\mu_{\text{NTP}}$ are the free energies of phosphodiester bond formation and NTP hydrolysis. \\

We use CFT's canonical coordinate $\mathbf{z}=(\boldsymbol{\pi}, \boldsymbol{\tau}, \mathfrak{F}_{\text{cycle}})$ to map these performance targets back to the underlying transition rates. Since each fundamental cycle is identified by a defining edge, or ``chord'', we swap the traffic variables on these chords for the fundamental cycle fluxes $\boldsymbol{J}_{\text{cycle}}$, leaving the traffics on all other edges as the remaining vector $\boldsymbol{\tau}_{\text{rem}}$. Expressing these cycle fluxes in terms of our target metrics establishes the functional coordinate $(\boldsymbol{\pi}, \boldsymbol{\tau}_{\text{rem}}, S, \varepsilon, C, r', \mathfrak{F}_{\text{cycle}})$, where $r' = J'_{\text{W}}/J'_{\text{R}}$ is the discard ratio. This yields an exact analytical map $k_{ij}(\boldsymbol{\pi}, \boldsymbol{\tau}_{\text{rem}}, S, \varepsilon, C, r', \mathfrak{F}_{\text{cycle}})$ that translates target performances and cycle force constraints directly into the required transition rate configurations (see SI Sec. III.A \cite{SI}).\\\\

This parameterization yields physical performance bounds dictated by the non-negativity of one-way transition fluxes ($p_{ij} = [\tau_{ij} + J_{ij}]/2 \ge 0$). For example, the kinetic capacity of the initial binding step of the Right substrate imposes the constraint $J_{\text{R}}(S,\varepsilon)+J'_{\text{R}} (S,C,r')\le \tau_{\text{E,ER}}$ (i.e., $p_{\text{ER} \to \text{E}}\ge 0$). This yields the inequality:
\begin{equation}
\label{eq: KP_bounds}
\frac{S}{1+10^\varepsilon} + \frac{C - S\Delta\mu_p}{\Delta\mu_{\text{NTP}}(1+r')} \le \tau_{\text{E,ER}}.
\end{equation}
This bound (numerically verified in Fig.~\ref{fig: KP inverse design}b) describes a traffic jam effect: productive incorporation (the first term) and futile proofreading (the second term) must compete for a finite kinetic bandwidth ($\tau_{\text{E,ER}}$). Because $S$, $\varepsilon$, and $C$ serve as independent coordinate dimensions, a trade-off manifests \textit{only} at the boundary where this inequality saturates. As long as the traffic bandwidth is not saturated (the feasible interior), these metrics are decoupled. An enzyme can simultaneously improve speed, reduce error, and lower cost under fixed thermodynamic cycle drives ($\mathfrak{F}_{\text{cycle}}$). Fig.~\ref{fig: KP inverse design}c demonstrates this by designing a linear trajectory that optimizes all three metrics from their wild-type origins. Whenever this designed path saturates a local traffic bound ($p_{ij} \to 0$, marked with dots), we iteratively relax that specific $\tau_{ij}$ limit tenfold, allowing the optimization to continue (see SI Sec. III.C for algorithmic details \cite{SI}).\\

At some point, simultaneously increasing speed and decreasing cost forces the proofreading fluxes to become negative ($J'_{\text{R}} + J'_{\text{W}} \le 0$). To sustain a lower cost, the system must synthesize NTP, entering an ``anti-proofreading'' regime (crossing the vertical dash-dotted line in Fig.~\ref{fig: KP inverse design}c, d) \footnote{We define ``anti-proofreading'' by the net reversal of the proofreading cycle, which operates via a different physical mechanism than the state-occupancy or speed-driven anti-proofreading regimes discussed previously by Murugan \textit{et al.} \cite{murugan_discriminatory_2014}.}. By feeding this targeted linear performance path back into our inverse map $k_{ij}(\boldsymbol{\pi}, \boldsymbol{\tau}_{\text{rem}}, S, \varepsilon, C, r', \mathfrak{F}_{\text{cycle}})$, we calculate the driving protocol of transition rates required to realize it (Fig.~\ref{fig: KP inverse design}d). Achieving this regime necessitates accelerating reversed transition rates (colored dashed lines in Fig.~\ref{fig: KP inverse design}d) to drive the proofreading cycle backwards. Because the parameters $\boldsymbol{\pi}$ and $r'$ remain freely tunable, the shown driving protocol is not unique---demonstrating a flexibility CFT uncovers for the design of kinetic proofreaders.\\

\begin{figure}
    \centering
    \includegraphics[width=\linewidth]{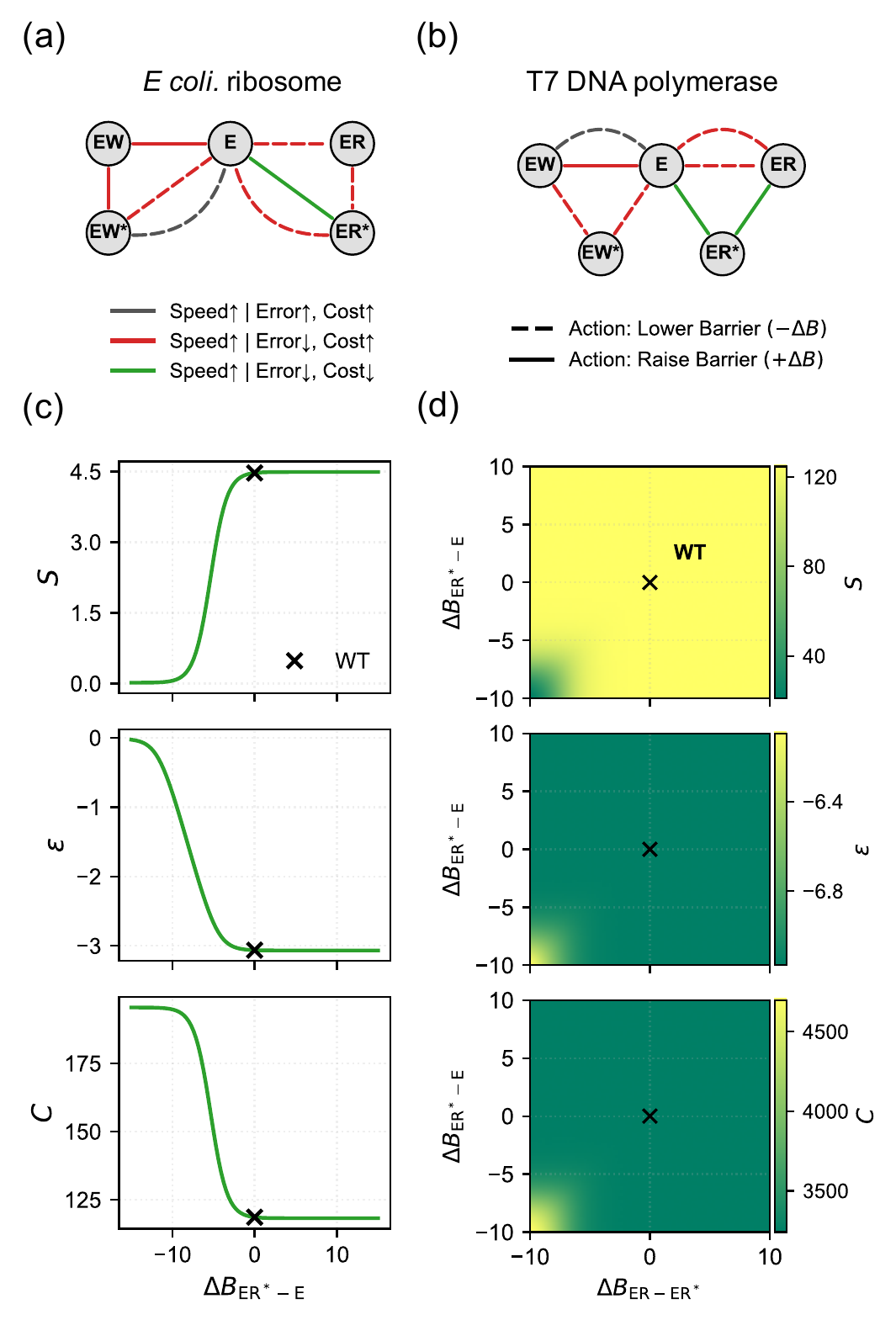}
    \caption{\textbf{Performance routing and evolutionary saturation in kinetic proofreading.}
    \textbf{(a, b)} Barrier sensitivity maps for T7 DNA polymerase (a) and \textit{E. coli} ribosome (b). Colors indicate the different types of responses for each barrier. Solid and dashed lines denote raising ($+\Delta B$) and lowering ($-\Delta B$) barriers to raise speed, respectively.
    \textbf{(c, d)} Performance metrics ($S, \varepsilon, C$) along the ``free lunch'' barrier tuning (Green in a,b). The wild-type (WT) parameters (marked as x) reside near the saturating plateau for all three metrics.  }
    \label{fig: KP barrier tuning}
\end{figure}

\textbf{Barrier routing reveals an evolutionary ``free lunch''.}
While the inversely designed protocol proves that decoupling speed, error, and cost is mathematically possible, applying CFT's response rules reveals a simpler physical implementation: this optimal routing can be achieved by tuning only specific kinetic barriers. Furthermore, evaluating the wild-type parameters demonstrates that biological evolution has already driven this specific tuning mechanism to near-saturation.\\

Using our analytical barrier routing rules (Eq.~\ref{eq: dx/dB}), we map the performance gradients for T7 DNA polymerase and the \textit{E. coli} ribosome (Fig.~\ref{fig: KP barrier tuning}a, b). This reveals three tuning directions. Most barrier perturbations enforce a standard trade-off (red): speed increases at the expense of higher energetic cost. Accelerating the Wrong incorporation step (gray) is detrimental, compounding both error and cost. Crucially, both networks share a topological ``free lunch'' direction (green) that simultaneously increases speed while lowering error and cost.\\

In both systems, this optimal protocol is accessed by raising the barrier in the discard pathway of the Right substrate. This theoretical requirement provides a physical rationale for the asymmetric rates observed in real proofreading networks \cite{banerjee_elucidating_2017,mallory_trade-offs_2019}---suggesting that evolution has developed structural mechanisms to selectively tune the cognate discard barrier without symmetrically affecting the non-cognate one.\\

Tracking the performance metrics along this specific path---the $\text{ER}^* \rightleftharpoons \text{E}$ transition in the ribosome, and the corresponding excision pathway in the polymerase---allows us to directly interrogate the wild-type enzymes. As illustrated in Fig.~\ref{fig: KP barrier tuning}c and d, the wild-type parameters reside near the plateau where the marginal gain of further raising this barrier almost vanishes. Biological evolution appears to have pushed this specific topological router to its optimal limit.\\

In summary, the canonical understanding of speed-accuracy-cost trade-offs in kinetic proofreading is boundary constraints \cite{murugan_speed_2012,sartori_thermodynamics_2015,rao_thermodynamics_2015,chiuchiu_pareto_2023}. Within the operational interior, these performance metrics can be decoupled and inversely designed.  This is consistent with the emerging realization that NEQ driving can circumvent static thermodynamic trade-offs \cite{goetz_non-equilibrium_2025}. Evolution appears to have navigated this interior region by pushing the cognate discard barrier along a ``free lunch'' tuning direction toward near-saturation. 

\section*{3) Stronger enzyme inhibition: Dead-ends versus leaky loops}
Classical pharmacology evaluates enzyme inhibitor efficacy based on binding affinities, $\ln (k_\text{on}/k_\text{off})$ of the inhibitor molecule to the target enzyme \cite{copeland_evaluation_2005}. By treating Michaelis-Menten (MM) inhibition as a NEQ flow-routing problem, we demonstrate that inhibitor efficacy fundamentally depends on network topology.  To analyze these networks, we must first address the non-invertible catalytic step ($\text{ES} \to \text{E} + \text{P}$). We extend CFT---originally formulated for networks where all transitions possess non-zero inverse rates---directly to this zero-rate limit. Because the catalytic transition is non-invertible, its symmetric traffic equals its directed flux ($\tau_{\text{cat}} = J_{\text{cat}}$). By selecting a spanning tree where this non-invertible edge acts as a fundamental cycle chord, maximizing the path entropy (Maximum Caliber) reveals that most conjugate forces preserve their original forms (see SI Sec. IV.A \cite{SI}). The single modification occurs on the catalytic cycle: the lack of an inverse transition alters its localized force contribution from $\frac{1}{2}\ln(k_{+}/k_{-})$ to $\ln k_{\text{cat}}$, yielding the cycle force:
\begin{equation}
    \mathfrak{F}_{\text{cat}} = \ln k_{\text{cat}} + \frac{1}{2} \ln \left(\frac{k_{\text{E,ES}}}{k_{\text{ES,E}}}\right).
\end{equation}
With this observable-force conjugacy established, CFT's analytical machinery remains valid for MM networks. We can now use it to evaluate the classical inhibition archetypes.\\

\begin{figure}
    \centering
    \includegraphics[width=0.75\linewidth]{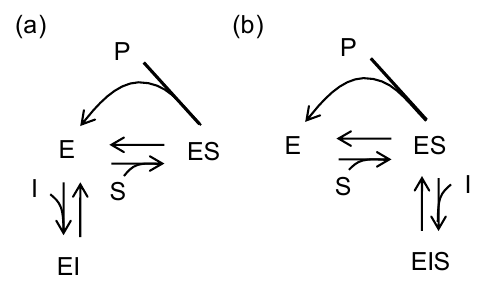}
    \caption{\textbf{Competitive and uncompetitive inhibitors rely on topological dead-ends.} Network diagrams for \textbf{(a)} competitive and \textbf{(b)} uncompetitive inhibition. Unlike the noncompetitive loop, the inhibitor-bound states (EI and EIS) form dead-end pathways carrying zero steady-state net flux. This topological distinction renders their long-term catalytic rates insensitive to kinetic barrier perturbations along the inhibitor pathways.}
    \label{fig: competitive and uncompetitive Enzyme Inhib}
\end{figure}

We first demonstrate that competitive and uncompetitive inhibitors act as topological dead-ends optimized solely by node energies (Fig.~\ref{fig: competitive and uncompetitive Enzyme Inhib}). The inhibitor-bound states (EI and EIS) form dead-end pathways that carry zero net flux ($J_{\text{sideway}} = 0$). According to CFT's barrier routing rule, tuning the kinetic barriers along these pathways has zero effect on the long-term catalytic production. Therefore, to optimize dead-end inhibitors, designers cannot rely on barrier routing; they must instead deepen the node energy ($E_n$) of the bound state, maximizing the binding affinity purely by suppressing $k_{\text{off}}$.\\

\begin{figure}
    \centering
    \includegraphics[width=0.8\linewidth]{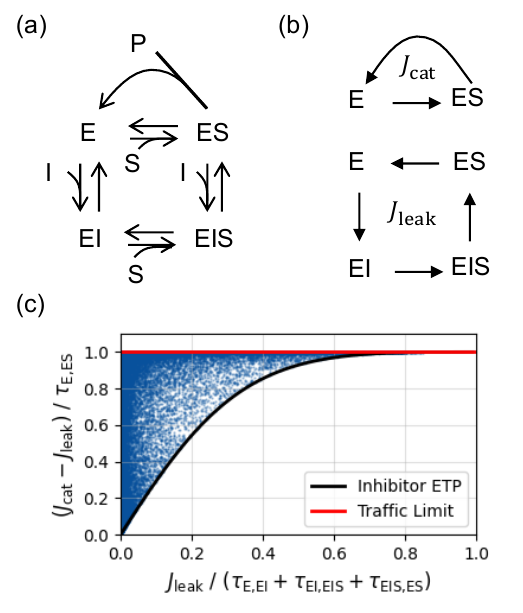}
    \caption{\textbf{Inhibitor Equal Traffic Principle in noncompetitive enzyme inhibition.} (a) The four-state network with a non-invertible catalytic step ($\text{ES} \to \text{E} + \text{P}$). (b) The two fundamental cycle fluxes: the non-invertible catalytic flux ($J_{\text{cat}}$) and the invertible leak flux ($J_{\text{leak}}$). (c) Numerical validation of the operational bounds. The blue scatter points represent $10^5$ randomly sampled kinetic networks with log-uniform transition rates spanning six orders of magnitude, constrained to maintain a futile inhibitor loop ($\mathfrak{F}_{\text{leak}}=0$). For a given $J_{\text{leak}}$, catalytic production is minimized when kinetic traffic is equally distributed across the sideway edges of the inhibitor loop (lower bound). Production is maximized when the reverse one-way flux to the free enzyme vanishes ($p_{\text{ES,E}} \to 0$, upper bound).}
    \label{fig: Non-comp MM Enzyme Inhib}
\end{figure}

In contrast, noncompetitive inhibitors form leaky loops that require kinetic barrier routing for optimization. These inhibitors form a closed loop ($\text{E} \rightleftharpoons \text{EI} \rightleftharpoons \text{EIS} \rightleftharpoons \text{ES} \rightleftharpoons \text{E}$) that couples to the catalytic step (Fig.~\ref{fig: Non-comp MM Enzyme Inhib}a). Although this internal futile cycle has zero cycle force ($\mathfrak{F}_{\text{leak}} = 0$), the non-invertible catalysis continually depletes $\text{ES}$ and replenishes $\text{E}$. This tilted population gradient acts as a pump, driving a compensatory clockwise leak flux ($J_{\text{leak}} > 0$) through the indirect inhibitor pathway (see SI Sec. IV.B for proof \cite{SI}). Because of this non-zero leak flux, the kinetic barriers act as effective control knobs for noncompetitive inhibition.\\

To optimize this network, we derive the \textit{Inhibitor Equal Traffic Principle} (iETP) as a multicircular generalization to ETP. Using the CFT coordinate $\mathbf{z}=(\boldsymbol{\pi}, \boldsymbol{\tau}, J_{\text{leak}}, \mathfrak{F}_{\text{leak}})$, we express the catalytic production \cite{SI}:
\begin{equation} \label{eq: leak_coupling}
    J_{\text{cat}} = J_{\text{leak}} + \tau_{\text{E,ES}} \tanh(\mathcal{A}_{\text{side}}),
\end{equation}
where $\mathcal{A}_{\text{side}} = \sum_{\eta} \text{arctanh}(J_{\text{leak}}/\tau_\eta)$ is the total affinity across the three sideway edges of the loop. Minimizing $J_{\text{cat}}$ under fixed traffic $\boldsymbol{\tau}$ requires minimizing $\mathcal{A}_{\text{side}}$, which dictates equalizing the traffic across these sideway edges. Therefore, maximizing inhibition requires a balanced traffic distribution across the leak loop. Conversely, the maximum catalytic production occurs at the traffic upper bound ($J_{\text{cat}}^{\max} = J_{\text{leak}} + \tau_{\text{E,ES}}$), where the reverse one-way flux $p_{\text{ES,E}}$ vanishes. We numerically validate these bounds in Fig.~\ref{fig: Non-comp MM Enzyme Inhib}c.\\

In summary, extending CFT to catalytic networks recasts enzyme inhibition from an EQ affinity problem into a NEQ topological flow-routing problem. Competitive and uncompetitive inhibitors act as topological dead-ends, blind to barrier routing and reliant entirely on node energy optimization. In contrast, noncompetitive inhibitors form leaky loops, reaching maximum efficacy only when kinetic barriers actively balance their internal traffic. This establishes a NEQ physical basis for targeted drug design.

\section*{Discussion}

This work reframes the design of nonequilibrium molecular machines as a flow-routing problem. Traditional thermodynamic models focus on energetic drives, net fluxes, and correlating machine performance with entropy production, but they lack explicit handles for the \textit{topological misflows} that degrade machine efficiency---such as futile backstepping in the F$_1$ motor, branched misrouting in kinetic proofreading, or dead-ends or leaks in enzyme inhibition. By pairing dynamic observables with conjugate path-entropic forces, Caliber Force Theory (CFT) establishes a dual design framework: \textit{inverse design} to map performance targets to optimal transition rate configurations, and \textit{forward design} to scale or route flows with node energies or kinetic barriers.\\

While we focused on biochemical networks governed by Local Detailed Balance and Arrhenius energetics, the framework is independent of these specific physical assumptions. Inherited from its path-entropy-based construction \cite{yang_principled_2026}, CFT Design applies to any ergodic Markov jump process and can be generalized to continuous processes in principle. This extends the applicability to other nonequilibrium systems, such as ecological population dynamics \cite{yang_nonequilibrium_2025} or vehicular traffic \cite{yang_principled_2026}. Just as equilibrium thermodynamics provides the foundational conjugate forces for static systems, CFT Design provides them for optimizing and controlling nonequilibrium flows.

\section*{Code Availability} 
The codes used to produce the Figures
are available at DOI: 10.5281/zenodo.20057619.\\

\begin{acknowledgments}
We are grateful for the financial support from the Laufer Center for Physical and Quantitative Biology at Stony Brook, the John Templeton Foundation (Grant ID 62564), and NIH (Grant RM1-GM135136).
\end{acknowledgments}

\section*{Author Contributions}
YJY and KD conceptualized the project together. YJY developed the
framework and the results. KD supervised all developments. Both authors contributed equally to writing and editing the manuscript.

\bibliography{Ref}

\clearpage
\onecolumngrid
\setcounter{equation}{0}
\setcounter{figure}{0}
\setcounter{table}{0}
\setcounter{page}{1}
\renewcommand{\theequation}{S\arabic{equation}}
\renewcommand{\thefigure}{S\arabic{figure}}
\renewcommand{\thetable}{S\arabic{table}}

\begin{center}
    {\large \textbf{Supplementary Information for:\\ ``Nonequilibrium Theory for Molecular Machine Design''}} \\
    
    \vspace{1em} 
    
    Ying-Jen Yang$^{1,*}$ and Ken A. Dill$^{1,2,3}$ \\
    
    \vspace{0.5em} 
    
    {\small \itshape
    $^{1}$Laufer Center of Physical and Quantitative Biology, Stony Brook University \\
    $^{2}$Department of Physics and Astronomy, Stony Brook University \\
    $^{3}$Department of Chemistry, Stony Brook University} \\
    
    \vspace{0.5em}
    
    {\small $^*$ying-jen.yang@stonybrook.edu} \\

    \vspace{1em} 

    (Dated: \today)
\end{center}

\starttocentries
\tableofcontents{}

\section{A Brief Review of Caliber Force Theory (CFT)}

The operational power of equilibrium (EQ) thermodynamics stems from its conjugate structure of extensive observables and intensive forces, formally linked by Legendre transforms. The origin of the conjugacy lies in the Boltzmann-Gibbs distribution as an exponential family maximizing state entropy (Max Ent). The resulting zoo of conjugate coordinates allows the handling of macroscopic constraints (e.g., isothermal, isobaric, isochoric) via ensemble transformations to the most suitable coordinate system (e.g., canonical, microcanonical). This mathematical structure exists because variables like $1/T$, $p/T$, and $-\mu/T$ are defined as entropic forces (derivatives of the state entropy potential), ensuring they are conjugate to the physical observables $U$, $V$, and $N$. Caliber Force Theory (CFT) is designed to mirror this exact conjugate structure for nonequilibrium (NEQ) flows within a network state space. In this section, we recapitulate the foundations of CFT, detailing the construction of conjugate coordinates via Maximum Caliber \cite{yang_principled_2026}, the derivation of the Response-Inverse-Matrix (RIM) relation \cite{yang_fluctuation-response_2026}, and the explicit proofs for the Arrhenius energetic perturbation rules utilized in the main text.

\subsection{Conjugate Coordinates and the Exponential Family}

Consider a continuous-time Markov jump process on a network state space consisting of $N$ nodes and $M$ edges, parameterized by transition rates $k_{ij}$ (from $i$ to $j$). We use the terminology that a transition is ``invertible'' if $k_{ij} > 0 \iff k_{ji} > 0$, and a system is ``reversible'' if it obeys detailed balance. We here assume all transitions are invertible (we will generalize this for non-invertible systems in Sec. IV), and the system can be either reversible or irreversible. The steady-state statistics of this system are fully described by the state probabilities $\pi_i$ and the one-way probability fluxes $p_{ij} = \pi_i k_{ij}$. Because elementary transitions $i\mapsto j$ are governed by independent Poisson processes, specifying these average observables entirely fixes the transition rates via $k_{ij} = p_{ij}/\pi_i$, thereby determining the full dynamics and all higher-order statistics.\\

To construct a ``microcanonical-like'' observable coordinate for dynamics, the chosen set of observables must be linearly independent. Raw transition counts and dwell times are degenerate due to probability conservation laws: probabilities must normalize ($\sum_i \pi_i = 1$), and steady-state fluxes must satisfy Kirchhoff's current law ($\sum_{j(\neq i)} p_{ij} = \sum_{j(\neq i)} p_{ji}$). CFT resolves this by projecting the dynamics onto a complete, non-degenerate observable basis. \\

Let $N_{ij}$ denote the number of $i \to j$ transitions and $T_n$ the dwell time in state $n$ over a trajectory of duration $t$. The fundamental counting observables are defined as $\mathbf{X} = (T_n, \Phi_{ij}, \Psi_c)$, where $\Phi_{ij} = N_{ij} + N_{ji}$ is the symmetric edge traffic count, and $\Psi_c = N_{ab} - N_{ba}$ is the antisymmetric net flux count across the defining chord $ab$ of fundamental cycle $c$. Their corresponding empirical rates are $\mathbf{x} = \mathbf{X}/t = (f_n, \phi_{ij}, \psi_c)$, whose long-term steady-state averages yield the observable basis $\bar{\textbf{x}} = (\pi_n, \tau_{ij}, J_c)$:
\begin{enumerate}
    \item \textbf{Node probabilities ($\pi_n$):} The state occupancies, yielding $N-1$ independent variables by excluding an arbitrary reference node $m$ to account for normalization.
    \item \textbf{Edge traffics ($\tau_{ij}$):} The symmetric, total kinetic activity on each edge, defined as $\tau_{ij} = p_{ij} + p_{ji}$, yielding $M$ variables (a term first coined by C. Maes and collaborators \cite{maes_frenesy_2020}).
    \item \textbf{Fundamental cycle fluxes ($J_c$):} The antisymmetric net fluxes $J_{ij} = p_{ij}-p_{ji}$ are degenerate due to Kirchhoff's current law. A standard method to find a linearly independent basis is to select a spanning tree of the network \cite{schnakenberg_network_1976,hill_free_1989}. Each chord (edge $ab$) added to the tree defines a fundamental cycle $c$, and its net flux $J_c = J_{ab} = \pi_a k_{ab} - \pi_b k_{ba}$ serves as an independent basis vector spanning all steady-state net flows. In network topology, a spanning tree leaves exactly $M - N + 1$ chords.
\end{enumerate}
Together, these form a complete, linearly independent ``microcanonical-like'' coordinate system of exactly $2M$ dynamical observables $\bar{\textbf{x}} = (\boldsymbol{\pi}, \boldsymbol{\tau}, \boldsymbol{J})$---comprising $(N-1) + M + (M-N+1) = 2M$ dimensions---which parameterize each rate as $k_{ij}(\bar{\textbf{x}})=[\tau_{ij}+J_{ij}(J_c)]/(2\pi_i)$.\\

\textbf{The Dimensionality Mismatch of Arrhenius Energetics:} 
To establish a formal framework for design, we seek a set of force variables strictly \textbf{conjugate} to these $2M$ observables, ensuring a one-to-one mapping and enabling coordinate swapability via Legendre transforms. One approach is to parameterize the rates using Arrhenius energetics: $k_{ij} = \mathcal{N}\exp[-(B_{ij}-E_i)]\exp(F_{ij}/2)$. However, for a connected network with $N$ nodes and $M$ edges, this parameterization utilizes $N$ node energies ($E_n$), $M$ kinetic barriers ($B_{ij}$), and $M$ edge affinities ($F_{ij}$), totaling $N + 2M$ energetic parameters.\\

Because the true kinetic degrees of freedom defining the Markov jump process are strictly $2M$, this fundamental dimensionality mismatch implies there are exactly $N$ degenerate dimensions in the Arrhenius parameterization. This gauge degeneracy can be demonstrated by introducing an arbitrary state potential $\phi_i$. If we shift the node energies by this potential ($E_i \to E_i + \phi_i$), we can perfectly absorb this shift and keep all transition rates $k_{ij}$ mathematically invariant by simultaneously adjusting the barriers and local affinities:
\begin{align}
    B_{ij} &\to B_{ij} + \frac{\phi_i + \phi_j}{2}, \\
    F_{ij} &\to F_{ij} + \phi_j - \phi_i.
\end{align}
Because $\sum_{(ij) \in c} (\phi_j - \phi_i) = 0$ around any closed loop, the thermodynamic cycle affinities ($\sum_c F_{ij}$) remain invariant under this gauge transformation. \\

This mathematical degeneracy proves that infinite combinations of $E$, $B$, and $F$ map to the exact same dynamics. Consequently, they cannot form a one-to-one conjugate coordinate system with the observables $\bar{\textbf{x}}$. Logically, this is expected: Arrhenius parameters describe specific energetic properties of a physical barrier-crossing model, whereas the observables $(\boldsymbol{\pi}, \boldsymbol{\tau}, \boldsymbol{J})$ describe the purely dynamical properties of the generic Markov process. One should not expect model-specific properties to serve as the generic dynamical conjugates. The true conjugate forces must emerge from the dynamics itself.\\

To establish a Legendre-dual parameterization for these dynamical observables and obtain the force coordinate variables, CFT employs the Principle of Maximum Caliber (Max Cal) \cite{presse_principles_2013,yang_principled_2026}, which is to NEQ dynamics what Max Ent is to EQ distributions. Let $\mathcal{P}_{\mathbf{k}}(\omega_t)$ be the probability of a path $\omega_t$ of duration $t$ generated by rates $\mathbf{k}$. Max Cal maximizes the relative path entropy rate, $\mathfrak{s}_{\text{path}} = \lim_{t\to\infty} -\frac{1}{t} \sum_{\omega_t} \mathcal{P}_{\mathbf{k}}(\omega_t) \ln[\mathcal{P}_{\mathbf{k}}(\omega_t)/\mathcal{P}_{\mathbf{u}}(\omega_t)]$, relative to a uniform reference ``doldrum'' process ($u_{ij}=1$), subject to constraints on the average observables $\bar{\textbf{x}}$. \\

This constrained optimization generates a dynamic \textit{exponential family} of path probabilities:
\begin{equation}
    \mathcal{P}_{\mathbf{k}}(\omega_t) = \mathcal{P}_{\mathbf{u}}(\omega_t) \frac{\exp \left( \boldsymbol{\mathfrak{F}} \cdot \mathbf{X}(\omega_t) \right)}{\mathcal{Z}_t(\boldsymbol{\mathfrak{F}})},
\end{equation}
where $\mathcal{Z}_t(\boldsymbol{\mathfrak{F}})$ is the dynamic partition function. A generic additive path observable takes the form $\int_0^t \mathbf{f}(i_s)ds + \sum_l \mathbf{g}(i_{t_l^-}, i_{t_l^+})$, where $s$ denotes continuous time and $l$ indexes the discrete jumps. The basis $\mathbf{X}$ is constructed by choosing specific indicator functions for $\mathbf{f}$ and $\mathbf{g}$: node dwell times use the Kronecker delta indicator $\mathbf{f}(x) = \delta_{x,n}$ (with $\mathbf{g}=0$); edge traffic counts use $\mathbf{g}(i,j) = \delta_{i,a}\delta_{j,b} + \delta_{i,b}\delta_{j,a}$ (with $\mathbf{f}=0$); and cycle net flux counts use $\mathbf{g}(i,j) = \delta_{i,a}\delta_{j,b} - \delta_{i,b}\delta_{j,a}$ on the defining chord $ab$.\\

The Lagrange multipliers $\boldsymbol{\mathfrak{F}}$ conjugate to these constraints are exactly the \textit{path entropic forces}. In the long-term limit, the scaled cumulant generating function---acting as the NEQ free energy or log dynamic partition function, defined as the Caliber rate $\mathfrak{c}(\boldsymbol{\mathfrak{F}}) = \lim_{t\to\infty} \frac{1}{t}\ln \mathcal{Z}_t(\boldsymbol{\mathfrak{F}})$---is obtained via the Legendre-Fenchel transform \cite{chetrite_nonequilibrium_2013}:
\begin{equation} \label{eq:SI_Legendre_Forward}
    \mathfrak{c}(\boldsymbol{\mathfrak{F}}) = \max_{\bar{\textbf{x}}} \left[ \mathfrak{s}_{\text{path}}(\bar{\textbf{x}}) + \boldsymbol{\mathfrak{F}} \cdot \bar{\textbf{x}} \right].
\end{equation}
This maximization explicitly defines the coordinate transformation from $\bar{\textbf{x}}$ to $\boldsymbol{\mathfrak{F}}$, guaranteeing that the forces are derivatives of the entropy: $\boldsymbol{\mathfrak{F}} = -\partial \mathfrak{s}_{\text{path}}/\partial \bar{\textbf{x}}$. Conversely, the inverse coordinate transform from $\boldsymbol{\mathfrak{F}}$ back to $\bar{\textbf{x}}$ is given by:
\begin{equation} \label{eq:SI_Legendre_Inverse}
    -\mathfrak{s}_{\text{path}}(\bar{\textbf{x}}) = \max_{\boldsymbol{\mathfrak{F}}} \left[ \boldsymbol{\mathfrak{F}} \cdot \bar{\textbf{x}} - \mathfrak{c}(\boldsymbol{\mathfrak{F}}) \right].
\end{equation}
This guarantees that the Caliber function is a generating function, $\bar{\textbf{x}} = \partial \mathfrak{c}/\partial \boldsymbol{\mathfrak{F}}$. 
The standard Legendre transform expression, $\mathfrak{c} = \mathfrak{s}_{\text{path}}(\bar{\textbf{x}}) + \boldsymbol{\mathfrak{F}} \cdot \bar{\textbf{x}}$, serves as a compressed summary of these two explicit optimizations: when evaluated in the $\boldsymbol{\mathfrak{F}}$ coordinate, it reflects Eq.~\eqref{eq:SI_Legendre_Forward}, and when evaluated in the $\bar{\textbf{x}}$ coordinate, it reflects Eq.~\eqref{eq:SI_Legendre_Inverse}.\\

A key advantage of this conjugate structure is the ability to construct other coordinates via partial Legendre transforms. For example, to parameterize the dynamics under a fixed cycle force $\mathfrak{F}_{\text{cycle}}$ while treating the internal kinetic layout (probabilities and traffics) as independent variables, one constructs the potential $\mathcal{F}(\pi_n, \tau_{ij}, \mathfrak{F}_{\text{cycle}})$:
\begin{equation}
    \mathcal{F}(\pi_n, \tau_{ij}, \mathfrak{F}_{\text{cycle}}) = \max_{{J}_c} \left[ \mathfrak{s}_{\text{path}}(\bar{\textbf{x}}) + {\mathfrak{F}}_{\text{cycle}} \cdot {J}_c \right].
\end{equation}
This directly mirrors the transformation from the microcanonical ensemble $(U,V,N)$ to the canonical one $(T,V,N)$ by the Helmholtz free energy $F(T,V,N)$ in equilibrium thermodynamics. The inverse partial Legendre-Fenchel transform then reveals how to compute the flux $J_c$ directly from the observable-force coordinate:
\begin{equation}
    -\mathfrak{s}_{\text{path}}(\bar{\textbf{x}}) = \max_{{\mathfrak{F}}_{\text{cycle}}} \left[ {\mathfrak{F}}_{\text{cycle}} \cdot{J}_c - \mathcal{F}(\pi_n, \tau_{ij}, \mathfrak{F}_{\text{cycle}}) \right],
\end{equation}
which guarantees $J_c = \partial \mathcal{F} / \partial \mathfrak{F}_{\text{cycle}}$. Other mixed coordinates can be constructed similarly. We will explicitly deploy this mixed-coordinate strategy in the subsequent sections to derive the Equal Traffic Principle (ETP) for molecular motors, the operational envelopes of kinetic proofreading, and the inhibitor generalizations of the ETP for enzyme inhibition.\\

To give these path entropic forces more concrete physical meanings, CFT solves for their transition rate expressions. CFT utilizes Large Deviation Theory (LDT) \cite{touchette_large_2009} to evaluate the Caliber via the eigenvalue problem of a ``tilted matrix'' $\tilde{\mathbf{M}}(\boldsymbol{\mathfrak{F}})$. The off-diagonal elements are tilted by the forces associated with the transitions, $\tilde{M}_{ij} = \exp(\boldsymbol{\mathfrak{F}}\cdot \mathbf{g}(i,j))$, while the diagonal elements capture node constraints, $\tilde{M}_{ii} = -N_i + \boldsymbol{\mathfrak{F}}\cdot \mathbf{f}(i)$, where $N_i$ is the out-degree of node $i$ (the number of connected neighbors) in the reference doldrum process. The largest eigenvalue of $\tilde{\mathbf{M}}$ is exactly the Caliber rate $\mathfrak{c}$, and the transition rates $k_{ij}$ are determined by the forces and the corresponding right eigenvectors $v_i$ of $\tilde{\mathbf{M}}$:
\begin{subequations}
\begin{align} \label{eq: k(F)}
    k_{ij}(\boldsymbol{\mathfrak{F}}) &= \frac{v_j(\boldsymbol{\mathfrak{F}})}{v_i(\boldsymbol{\mathfrak{F}})} e^{\boldsymbol{\mathfrak{F}} \cdot \mathbf{g}(i,j)} \quad (i \neq j), \quad \text{and}\\
    \quad k_{ii}(\boldsymbol{\mathfrak{F}}) &= -N_i + \boldsymbol{\mathfrak{F}}\cdot \mathbf{f}(i) - \mathfrak{c}(\boldsymbol{\mathfrak{F}}).
\end{align}
\end{subequations}
These relations map the forces back to the transition rates.\\

By taking the logarithmic ratios of these transition rates along specific topological structures---such as the product of forward and backward rates on an edge, or the product of rates around a closed loop---the eigenvector ratios $v_j/v_i$ perfectly telescope and cancel out. This algebraic cancellation yields the explicit expressions for the conjugate forces $\boldsymbol{\mathfrak{F}} = (\mathfrak{F}_{\text{node},n}, \mathfrak{F}_{\text{edge},ij}, \mathfrak{F}_{\text{cycle},c})$ directly in terms of the rates \cite{yang_principled_2026}:
\begin{align}
    \mathfrak{F}_{\text{node},n} &= \sum_{i(\neq m)}(k_{mi}-1)-\sum_{j(\neq n)}(k_{nj}-1), \label{eq:SI_force_node} \\
    \mathfrak{F}_{\text{edge},ij} &= \frac{1}{2}\ln{k_{ij}k_{ji}}, \label{eq:SI_force_edge} \\
    \mathfrak{F}_{\text{cycle},c} &= \frac{1}{2}\sum_{(ij)\in c}\ln\frac{k_{ij}}{k_{ji}}. \label{eq:SI_force_cycle}
\end{align}
Deriving these explicit rate expressions is a critical theoretical step. It grounds the path entropic forces with clear physical interpretations: they are the specific affinities for node escaping, edge exchange, and cycle completion. Practically, connecting the forces directly to the transition rates makes the abstract concept of a ``fixed force constraint'' physically realizable. This ultimately allows us to tackle explicit force constraints and evaluate network responses to kinetic rate or energetic perturbations.

\subsection{General Response Relations and Symmetries}

Because the Caliber $\mathfrak{c}(\boldsymbol{\mathfrak{F}})$ is the scaled cumulant generating function, its derivatives directly yield the fluctuation-response relations. Specifically, the sensitivity of an average observable to its conjugate force evaluates to their asymptotic covariance:
\begin{equation}
    \frac{\partial \bar{x}_\alpha }{\partial \mathfrak{F}_\beta} = \lim_{t\to\infty} t \, \mathrm{Cov}(x_\alpha, x_\beta) = \mathbf{C}_{\alpha\beta},
\end{equation}
when \textit{all other forces are held fixed}. Here, the indices $\alpha$ and $\beta$ run through the path-dependent, stochastic observables $\mathbf{x} = (f_n, \phi_{ij}, \psi_c)$ defined in the previous section.\\

However, tuning forces while clamping other forces is not the same as tuning rates while fixing other rates. To map the fundamental force-response relations to the rate-response behaviors, we evaluate the sensitivity using the chain rule:
\begin{equation}
    k_{ij}\frac{\partial \bar{x}_\alpha}{\partial k_{ij}} = \sum_\beta \frac{\partial \bar{x}_\alpha}{\partial \mathfrak{F}_\beta} k_{ij}\frac{\partial \mathfrak{F}_\beta}{\partial k_{ij}} = \sum_\beta \mathbf{C}_{\alpha\beta} \mathbf{A}_{(ij),\beta}.
\end{equation}
This naturally introduces the sparse Jacobian matrix $\mathbf{A}$, defined as:
\begin{equation}
    \mathbf{A}_{(ij),\beta} \equiv k_{ij} \frac{\partial \mathfrak{F}_\beta}{\partial k_{ij}} = \frac{\partial \mathfrak{F}_\beta}{\partial \ln k_{ij}}.
\end{equation}
Based on the explicit force expressions (Eqs. \ref{eq:SI_force_node}-\ref{eq:SI_force_cycle}), the structure of $\mathbf{A}$ strictly encodes the local node-edge-cycle topology of the network, which will be useful in identifying response symmetries valid for any network topology.\\

To evaluate this rate sensitivity $k_{ij}\partial_{k_{ij}} \bar{x}_\alpha$, we parse the covariance matrix $\mathbf{C}_{\alpha\beta}$ using the intrinsic stochasticity of the system \cite{yang_fluctuation-response_2026}. In a Markov jump process, the independent Poissonian noise source on each transition edge is defined as $\lambda_{ij} = (N_{ij} - T_i k_{ij})/t$. Because the raw counts $N$ and $T$ can be linearly spanned by the fundamental observables $\mathbf{X}$, the noise sources $\boldsymbol{\lambda}$ can be asymptotically expanded as a linear combination of the empirical rates $\mathbf{x} = \mathbf{X}/t$. It turns out this linear mapping is governed by the exact same Jacobian matrix $\mathbf{A}$ \cite{yang_fluctuation-response_2026}:
\begin{equation}
    \boldsymbol{\lambda} \sim \mathbf{A}\mathbf{x} - \nabla_{\ln \mathbf{k}} \mathfrak{c}.
\end{equation}
The term $\nabla_{\ln \mathbf{k}} \mathfrak{c}$ is simply a deterministic shift. Because the Caliber equates to the escape rate of the reference node $m$ up to a constant (i.e., $\mathfrak{c} = \sum_{j(\neq m)} (k_{mj} - 1)$), the derivative $k_{uv}\partial_{k_{uv}}\mathfrak{c}$ is non-zero (yielding $k_{mj}$) only when the perturbed edge originates from the reference node ($u=m, v=j$). \\

Inverting this linear mapping yields $\mathbf{x} \sim \mathbf{A}^{-1}(\boldsymbol{\lambda} + \nabla_{\ln \mathbf{k}} \mathfrak{c})$. In the long-time limit, the independent noise sources possess a diagonal covariance: $\lim_{t\to\infty} t \, \mathrm{Cov}(\lambda_{ij}, \lambda_{ab}) = \pi_i k_{ij} \delta_{ia}\delta_{jb}$. Consequently, the covariance matrix of the observables is completely spanned by the independent edge noises via the inverse Jacobian:
\begin{equation} \label{eq: cov decomposition}
    \mathbf{C}_{\alpha\beta} = \sum_{ab} [\mathbf{A}^{-1}]_{\alpha, (ab)} [\mathbf{A}^{-1}]_{\beta, (ab)} (\pi_a k_{ab}).
\end{equation}
Substituting this covariance decomposition back into our chain rule expression yields:
\begin{equation}
    k_{ij} \frac{\partial \bar{x}_\alpha}{\partial k_{ij}} = \sum_\beta \sum_{ab} [\mathbf{A}^{-1}]_{\alpha, (ab)} [\mathbf{A}^{-1}]_{\beta, (ab)} (\pi_a k_{ab}) \mathbf{A}_{(ij),\beta}.
\end{equation}
Because $\sum_\beta [\mathbf{A}^{-1}]_{\beta, (ab)} \mathbf{A}_{(ij),\beta} = \delta_{(ij),(ab)}$, the summation over $ab$ collapses. This algebraic simplification perfectly cancels out one of the inverse matrices, yielding the \textbf{Response-Inverse-Matrix (RIM) relation} (first derived in \cite{yang_fluctuation-response_2026}):
\begin{equation}
    k_{ij} \frac{\partial \bar{x}_\alpha}{\partial k_{ij}} = p_{ij} [\mathbf{A}^{-1}]_{\alpha, (ij)}. \label{eq:SI_RIM}
\end{equation}
\\

The algebraic identity $\mathbf{A}^{-1}\mathbf{A} = \mathbf{I}$ generates a set of universal response symmetries. Multiplying a row $\alpha$ of $\mathbf{A}^{-1}$ (corresponding to an observable $\bar{x}_\alpha$) by a column $\beta$ of $\mathbf{A}$ (corresponding to the conjugate force $\mathfrak{F}_\beta$) yields $\sum_{ij} [\mathbf{A}^{-1}]_{\alpha, (ij)} \mathbf{A}_{(ij), \beta} = \delta_{\alpha, \beta}$. Substituting the RIM relation translates these abstract matrix identities into three strict physical constraints on rate sensitivities \cite{yang_fluctuation-response_2026}:
\begin{enumerate}
    \item \textbf{Node Escaping Symmetry} (from column $\beta = f_n$ corresponding to $\mathfrak{F}_{\text{node},n}$):
    \begin{equation} \label{eq:SI_Node_Symmetry}
        \sum_{l(\neq m)}\frac{k_{ml}}{\pi_{m}}\frac{\partial \bar{x}_\alpha}{\partial k_{ml}}-\sum_{j(\neq n)}\frac{k_{nj}}{\pi_{n}}\frac{\partial \bar{x}_\alpha}{\partial k_{nj}}= \delta_{\bar{x}_\alpha,\pi_n}.
    \end{equation}
    \item \textbf{Edge Reciprocity} (from column $\beta = \phi_{ij}$ corresponding to $\mathfrak{F}_{\text{edge},ij}$):
    \begin{equation} \label{eq:SI_Edge_Reciprocity}
        \frac{1}{\pi_i}\frac{\partial \bar{x}_\alpha}{\partial k_{ij}} + \frac{1}{\pi_j}\frac{\partial \bar{x}_\alpha}{\partial k_{ji}} = 2 \delta_{\bar{x}_\alpha, \tau_{ij}}.
    \end{equation}
    \item \textbf{Cycle Symmetry} (from column $\beta = \psi_c$ corresponding to $\mathfrak{F}_{\text{cycle},c}$):
    \begin{equation} \label{eq:SI_Cycle_Symmetry}
        \sum_{(ij)\in c^{+}}\frac{1}{\pi_{i}}\frac{\partial \bar{x}_\alpha}{\partial k_{ij}}-\sum_{(ji)\in c^{-}}\frac{1}{\pi_{j}}\frac{\partial \bar{x}_\alpha}{\partial k_{ji}}=2 \delta_{\bar{x}_\alpha,J_c}.
    \end{equation}
\end{enumerate}

These fundamental topological symmetries dictate the network's steady-state response to specific classes of kinetic parameter perturbations. In the Arrhenius kinetic parameterization \cite{owen_universal_2020}, $k_{ij} \propto \exp[-(B_{ij}-E_i)]\exp(F_{ij}/2)$. This parameterization naturally classifies perturbations into state energy ($E$), kinetic barrier ($B$), and driving force ($F$) tuning. We explicitly evaluate the responses to state energy and kinetic barrier tuning below as these are the tunable knobs for molecular machines under fixed cycle forces.

\subsubsection{Node Energy as an Isotropic Scaler}
Perturbing the energy $E_n$ of state $n$ modifies all outgoing transition rates from that state proportionally. The perturbation operator is $\partial_{E_n} \equiv \sum_{j(\neq n)} k_{nj} \partial_{k_{nj}}$. Substituting this operator directly into the Node Escaping Symmetry (Eq. \ref{eq:SI_Node_Symmetry}) yields:
\begin{equation}
    \frac{1}{\pi_m}\frac{\partial \bar{x}_\alpha}{\partial E_m} - \frac{1}{\pi_n}\frac{\partial \bar{x}_\alpha}{\partial E_n} = \delta_{\bar{x}_\alpha,\pi_n}.
\end{equation}
For any flow observable distinct from probabilities (e.g., $\bar{x}_\alpha \in \{\tau_{ab}, J_c\}$), the Kronecker delta is zero. This requires that the normalized sensitivity $\frac{1}{\pi_n} \partial_{E_n} \bar{x}_\alpha$ is a global constant identical across all nodes in the network. Because steady-state fluxes and traffics possess physical units of inverse time, they scale linearly with a uniform, global shift in all node energies (which homogeneously scales all transition rates). By Euler's homogeneous function theorem of degree 1 \cite{yang_fluctuation-response_2026}, summing over all node perturbations yields $\sum_n \partial_{E_n} \bar{x}_\alpha = \bar{x}_\alpha$. Therefore, this network-wide constant must be $\bar{x}_\alpha$ itself, resulting in:
\begin{equation}
    \frac{\partial \bar{x}_\alpha}{\partial E_n} = \pi_n \bar{x}_\alpha.
\end{equation}
This shows that node energies act purely as isotropic scalers. They scale the global magnitude of all network flows proportionally but cannot alter flux ratios.

\subsubsection{Kinetic Barrier as a Topological Router}
Perturbing the symmetric kinetic barrier $B_{ij}$ on an edge modifies both the forward and backward rates equally and oppositely in logarithmic space. The operator is $\partial_{B_{ij}} \equiv -k_{ij}\partial_{k_{ij}} - k_{ji}\partial_{k_{ji}}$. 
Applying the Edge Reciprocity symmetry (Eq. \ref{eq:SI_Edge_Reciprocity}), we substitute $k_{ji}\partial_{k_{ji}}\bar{x}_\alpha = 2\pi_j k_{ji} \delta_{\bar{x}_\alpha, \tau_{ij}} - \frac{\pi_j k_{ji}}{\pi_i} \partial_{k_{ij}}\bar{x}_\alpha$ directly into the barrier operator. This yields:
\begin{align}
    \frac{\partial \bar{x}_\alpha}{\partial B_{ij}} &= -k_{ij}\frac{\partial \bar{x}_\alpha}{\partial k_{ij}} - \left( 2\pi_j k_{ji} \delta_{\bar{x}_\alpha, \tau_{ij}} - \frac{\pi_j k_{ji}}{\pi_i} \frac{\partial \bar{x}_\alpha}{\partial k_{ij}} \right) \nonumber \\
    &= -\frac{1}{\pi_i} \underbrace{\left( \pi_i k_{ij} - \pi_j k_{ji} \right)}_{J_{ij}} \frac{\partial \bar{x}_\alpha}{\partial k_{ij}} - 2 p_{ji} \delta_{\bar{x}_\alpha, \tau_{ij}}. \label{eq:SI_Barrier_Router_General}
\end{align}
This response formula branches into two distinct cases depending on the target observable:
\begin{enumerate}
    \item \textbf{For the local traffic ($\bar{x}_\alpha = \tau_{ij}$):} The Kronecker delta is $1$, yielding:
    \begin{equation}
        \frac{\partial \tau_{ij}}{\partial B_{ij}} = -\frac{J_{ij}}{\pi_i} \frac{\partial \tau_{ij}}{\partial k_{ij}} - 2 p_{ji}.
    \end{equation}
    Even if the edge is detailed-balanced ($J_{ij} = 0$), raising the barrier still strictly reduces the local kinetic activity (traffic) by $-2p_{ji}$.
    \item \textbf{For all other observables ($\bar{x}_\alpha \neq \tau_{ij}$):} The Kronecker delta vanishes, yielding:
    \begin{equation}
        \frac{\partial \bar{x}_\alpha}{\partial B_{ij}} = -\frac{J_{ij}}{\pi_i} \frac{\partial \bar{x}_\alpha}{\partial k_{ij}} = -J_{ij} [\mathbf{A}^{-1}]_{\alpha,(ij)}.
    \end{equation}
    If the perturbed edge carries a zero net flux ($J_{ij} = 0$), the steady-state response $\partial_{B_{ij}} \bar{x}_\alpha$ is exactly zero. This explicitly establishes kinetic barriers as the exclusive routers of the network: a barrier perturbation can only influence the global distribution or flux if it is placed on an edge carrying a non-zero nonequilibrium net flux.
\end{enumerate}

We remark that the Arrhenius energetic perturbations ($E, B$) discussed here are fundamentally distinct from tuning the intrinsic CFT forces ($\mathfrak{F}_{\text{node}}, \mathfrak{F}_{\text{edge}}$). Modifying a node energy $E$ or a kinetic barrier $B$ alters multiple transition rates simultaneously. In the CFT coordinate space, this corresponds to shifting several node and edge forces concurrently (i.e., moving vertically within the phase space where cycle forces are held constant). Nevertheless, their resulting responses are elegantly constrained by the topological symmetries derived via the Jacobian matrix. The theories about response relations presented up to this point---including the Jacobian mapping $\mathbf{A}$, the RIM relation, and the universal response symmetries---were a recapitulation of Ref.~\cite{yang_fluctuation-response_2026}. In later sections, we will utilize these established tools and symmetries to discover new operational principles for molecular machines.

\section{Example 1: F$_1$-ATPase Molecular Rotors}

\subsection{The Gerritsma and Gaspard (2010) model and implementation details}

To quantify the performance and operational bounds of the F$_1$-ATPase \textit{in vitro}, we adopt the two-state, two-edge Markov network model parameterized by Gerritsma and Gaspard (2010) \cite{gerritsma_chemomechanical_2010}. In this minimal model, the rotor transitions between an empty state (node 0) and a nucleotide-occupied state (node 1). These states are connected by two composite edges: the edge (T) representing ATP binding and unbinding, and the edge (D) representing ADP binding and release.\\

The four transition rates governing the network dynamics are defined as:
\begin{align}
    k_{\text{T}+}(\Gamma) &= k_{+1}(\Gamma, \zeta) \, [\text{ATP}] \\
    k_{\text{T}-}(\Gamma) &= k_{-1}(\Gamma, \zeta) \\
    k_{\text{D}-}(\Gamma) &= k_{+2}(\Gamma, \zeta) \\
    k_{\text{D}+}(\Gamma) &= k_{-2}(\Gamma, \zeta) \, [\text{ADP}][\text{P}_i]
\end{align}
where $\Gamma$ is the applied external torque and $\zeta$ is the viscous friction coefficient of the attached probe. \\

Based on their coarse-graining of a continuous Langevin model, the base rate constants $k_\rho$ ($\rho \in \{+1, +2, -2\}$) incorporate the friction dependence via a non-Arrhenius crossover form:
\begin{equation} \label{eq:SI_GG_rate_form}
    k_\rho(\Gamma, \zeta) = \frac{1}{e^{a_\rho(\Gamma)} + \zeta e^{b_\rho(\Gamma)}}
\end{equation}
The torque-dependent coefficients are expanded up to the second order: $y_\rho(\Gamma) = y_\rho^{(0)} + y_\rho^{(1)}\Gamma + y_\rho^{(2)}\Gamma^2$ (for $y \in \{a, b\}$). The specific values for these coefficients, which ensure the model accurately reproduces the single-molecule experiments (see references within \cite{gerritsma_chemomechanical_2010}), are summarized in Table \ref{tab:GG2010_params}.\\

The remaining ATP unbinding rate $k_{-1}$ is determined by the thermodynamic local detailed balance (LDB) condition along the cycle:
\begin{equation}
    k_{-1}(\Gamma, \zeta) = \frac{k_{+1}(\Gamma, \zeta)k_{+2}(\Gamma, \zeta)}{k_{-2}(\Gamma, \zeta)} \exp \left( \frac{\Delta G^0 - \frac{2\pi}{3}\Gamma}{k_B T} \right)
\end{equation}
where $\Delta G^0 = -50$ pN nm is the standard free energy of ATP hydrolysis.\\

\begin{table}[]
\centering
\caption{\textbf{Coefficients for the F$_1$-ATPase transition rates.} Parameters are adapted from Gerritsma and Gaspard (2010) \cite{gerritsma_chemomechanical_2010}.}
\label{tab:GG2010_params}
\begin{tabular}{lcccc}
\hline \hline
Coefficient & $k_{+1}$ (ATP binding) & $k_{+2}$ (Product release) & $k_{-2}$ (Product binding) \\ \hline
$a^{(0)}$ & -16.952 & -5.973 & -19.382 \\
$a^{(1)}$ (pN$^{-1}$ nm$^{-1}$) & $9.8 \times 10^{-4}$ & $1.7 \times 10^{-4}$ & $1.29 \times 10^{-1}$ \\
$a^{(2)}$ (pN$^{-2}$ nm$^{-2}$) & $5.8 \times 10^{-4}$ & $1.0 \times 10^{-3}$ & $2.8 \times 10^{-4}$ \\ \hline
$b^{(0)}$ & -16.352 & -2.960 & -18.338 \\
$b^{(1)}$ (pN$^{-1}$ nm$^{-1}$) & $-2.7 \times 10^{-2}$ & $-6.6 \times 10^{-4}$ & $5.9 \times 10^{-3}$ \\
$b^{(2)}$ (pN$^{-2}$ nm$^{-2}$) & $1.0 \times 10^{-3}$ & $3.6 \times 10^{-4}$ & $-2.1 \times 10^{-4}$ \\ \hline \hline
\end{tabular}
\end{table}

\textbf{Operational Regime and Simulation Parameters:} 
To maintain the physical validity of the discrete-state coarse-graining and to avoid mechanical slipping---which would violate the tight-coupling assumption inherent to the 2-state network---we explicitly restrict our analysis to the applied torque range of $\Gamma \in [-30, 30]$ pN nm, as specified in the original study \cite{gerritsma_chemomechanical_2010}. \\

For all numerical evaluations and figures presented in this work, the thermal energy is set to $k_{\text{B}}T = 4.1$ pN nm. The viscous friction coefficient $\zeta$ corresponds to a spherical probe of diameter $d = 160$ nm, evaluated as $\zeta = 2\pi \times 10^{-9} \times (d/2)^3 \times 4.75 \approx 0.0153$ pN$\cdot$nm$\cdot$s. Furthermore, we simulate the \textit{in vitro} low-ATP conditions using the specific nucleotide concentrations: $[\text{ATP}] = 0.49\ \mu$M, $[\text{ADP}] = 100\ \mu$M, and $[\text{P}_i] = 1$ mM. Under these exact boundary conditions, the total chemical driving force evaluates to $\Delta\mu = -\Delta G^0 + k_B T \ln([\text{ATP}] / [\text{ADP}][\text{P}_i]) \approx 56.5$ pN nm. This explicitly dictates the thermodynamic stalling torque where the cycle flux vanishes, $\Gamma_{\text{stall}} = - \frac{3}{2\pi} \Delta\mu \approx -27.0$ pN nm. Finally, owing to this low ATP concentration, the ATP edge (T) acts as the dominant kinetic bottleneck throughout the entire functional cycle.

\subsection{Equal Traffic processivity bound and inverse design}

As established in the main text (Tool 1), Caliber Force Theory (CFT) permits switching between independent control variables, establishing the observable-force ``canonical-like'' coordinate $\mathbf{z}=(\boldsymbol{\pi}, \boldsymbol{\tau}, \boldsymbol{\mathfrak{F}}_{\text{cycle}})$. This framework isolates the macroscopic free-energy drive (cycle force) from the internal kinetic layout. The associated dynamic free energy is the Legendre transform of the path entropy:
\begin{equation} 
\mathcal{F}(\mathbf{z}) = \sum_c J_c~\mathfrak{F}_{\text{cycle},c} + \mathfrak{s}_{\text{path}}, 
\end{equation} 
whose differentiation directly recovers the cycle flux: $J_c (\mathbf{z}) = \partial \mathcal{F}(\mathbf{z}) / \partial \mathfrak{F}_{\text{cycle},c}$. For the two-edge F$_1$-ATPase network, this separation allows us to perform systematic inverse design by explicitly solving for the cycle flux $J_c$ as a function of the traffic distribution.\\

The cycle force constraint dictates that the total environmental drive is exactly the sum of the internal edge affinities. In terms of the observable coordinates, this imposes the constraint:
\begin{equation} \label{eq:SI_arctanh_constraint}
    \mathfrak{F}_{\text{cycle}} = \sum_{i=1}^N \text{arctanh}\left(\frac{J_c}{\tau_i}\right).
\end{equation}
Before deriving the exact solution for the F$_1$-ATPase, we briefly recap the general \textit{Equal Traffic Principle} (ETP) bound for an $N$-step cycle \cite{yang_principled_2026}. Because the function $\text{arctanh}(J_c/\tau)$ is strictly convex with respect to $\tau$ (for $\tau > |J_c|$), applying Jensen's inequality to Eq.~\eqref{eq:SI_arctanh_constraint} under a fixed total kinetic activity budget ($\tau_{\text{tot}} = \sum \tau_i$) yields:
\begin{equation}
    \frac{\mathfrak{F}_{\text{cycle}}}{N} = \frac{1}{N}\sum_{i=1}^N \text{arctanh}\left(\frac{J_c}{\tau_i}\right) \ge \text{arctanh}\left( \frac{J_c}{\frac{1}{N}\tau_{\text{tot}}} \right).
\end{equation}
Rearranging this inequality establishes the upper bound on the machine's processivity \cite{yang_principled_2026}:
\begin{equation} \label{eq:SI_N_step_bound}
    \frac{J_c}{\tau_{\text{tot}}} \le \frac{1}{N} \tanh\left( \frac{\mathfrak{F}_{\text{cycle}}}{N} \right)
\end{equation}
which saturates at equal traffic distributed across the cycle.\\

\textbf{Exact Analytical Solution ($N=2$):} For the 2-edge F$_1$-ATPase network, we can solve the constraint exactly. Using the traffic fractions $\tau_\text{T} = \rho \tau_{\text{tot}}$ and $\tau_\text{D} = (1-\rho) \tau_{\text{tot}}$, where $\rho \in (0,1)$ is the traffic distribution ratio, Eq.~\eqref{eq:SI_arctanh_constraint} becomes $\mathfrak{F}_{\text{cycle}} = \text{arctanh}(J_c/\tau_\text{T}) + \text{arctanh}(J_c/\tau_\text{D})$. Applying the hyperbolic tangent addition formula, $\text{arctanh}(a) + \text{arctanh}(b) = \text{arctanh}[(a+b)/(1+ab)]$, we rewrite this as:
\begin{equation}
    \tanh(\mathfrak{F}_{\text{cycle}}) = \frac{\frac{J_c}{\tau_\text{T}} + \frac{J_c}{\tau_\text{D}}}{1 + \frac{J_c^2}{\tau_\text{T} \tau_\text{D}}} = \frac{J_c (\tau_\text{T} + \tau_\text{D}) \tau_\text{T} \tau_\text{D}}{\tau_\text{T} \tau_\text{D} (\tau_\text{T} \tau_\text{D} + J_c^2)} = \frac{J_c \tau_{\text{tot}}}{\tau_\text{T} \tau_\text{D} + J_c^2}.
\end{equation}
Rearranging this yields a quadratic equation for the cycle flux $J_c$:
\begin{equation}
    J_c^2 - \big[\tau_{\text{tot}} \coth(\mathfrak{F}_{\text{cycle}})\big] J_c + \tau_\text{T} \tau_\text{D} = 0.
\end{equation}
Because the physical net flux cannot exceed the symmetric traffic on any edge ($|J_c| \le \min(\tau_\text{T}, \tau_\text{D})$), the physically valid root is strictly the smaller one. Substituting $\tau_\text{T}\tau_\text{D} = \rho(1-\rho)\tau_{\text{tot}}^2$, we obtain the exact analytical mapping from the mixed coordinates to the cycle flux:
\begin{equation} \label{eq:SI_J_analytical}
    J_c(\rho, \tau_{\text{tot}}, \mathfrak{F}_{\text{cycle}}) = \frac{\tau_{\text{tot}}}{2} \left[ \coth(\mathfrak{F}_{\text{cycle}}) - \sqrt{\coth^2(\mathfrak{F}_{\text{cycle}}) - 4\rho(1-\rho)} \right].
\end{equation}
From a formal CFT perspective, obtaining this explicit algebraic solution is mathematically equivalent to computing the exact derivative of the free energy potential, $J_c = \partial \mathcal{F} / \partial \mathfrak{F}_{\text{cycle}}$.\\

\textbf{Processivity Bound (ETP):} The analytical solution Eq.~\eqref{eq:SI_J_analytical} makes the Equal Traffic Principle (ETP) strictly evident. For a fixed total kinetic activity $\tau_{\text{tot}}$ and fixed driving force $\mathfrak{F}_{\text{cycle}}$, the flux $J_c$ is maximized when the term inside the square root is minimized. This occurs exactly at the symmetric vertex $\rho = 1/2$ (i.e., $\tau_\text{T} = \tau_\text{D} = \tau_{\text{tot}}/2$). At this ETP optimum, the term $4\rho(1-\rho) = 1$, and utilizing the identity $\coth^2(x) - 1 = \text{csch}^2(x)$, the maximum achievable flux simplifies to:
\begin{equation}
    \max (J_c) = \frac{\tau_{\text{tot}}}{2} \left[ \frac{\cosh(\mathfrak{F}_{\text{cycle}}) - 1}{\sinh(\mathfrak{F}_{\text{cycle}})} \right].
\end{equation}
Applying the half-angle identities $\cosh(x)-1 = 2\sinh^2(x/2)$ and $\sinh(x) = 2\sinh(x/2)\cosh(x/2)$, the processivity bound rigorously simplifies to:
\begin{equation}
    \max \left( \frac{J_c}{\tau_{\text{tot}}} \right) = \frac{1}{2} \tanh\left(\frac{\mathfrak{F}_{\text{cycle}}}{2}\right).
\end{equation}
This proves the exact processivity bound, perfectly matching the $N=2$ case of the general inequality (Eq.~\ref{eq:SI_N_step_bound}).\\

\textbf{Rate Reconstruction and $\boldsymbol{\pi}$-Degeneracy:} 
Once $J_c$ is computed via Eq.~\eqref{eq:SI_J_analytical} for any chosen traffic distribution $\rho$, the complete set of transition rates can be unambiguously reconstructed for any arbitrary state distribution $\pi_0 \in (0,1)$ (with $\pi_1 = 1-\pi_0$). Utilizing the fundamental relations $k_{ij} = [\tau_{ij} + J_{ij}(J_c)] / (2\pi_i)$, the inverse design mapping is explicitly given by:
\begin{align} \label{eq:SI_rate_reconstruction}
    k_{\text{T}+} &= \frac{\rho \tau_{\text{tot}} + J_c}{2\pi_0} \quad & \quad k_{\text{T}-} &= \frac{\rho \tau_{\text{tot}} - J_c}{2\pi_1} \nonumber \\
    k_{\text{D}-} &= \frac{(1-\rho) \tau_{\text{tot}} + J_c}{2\pi_1} \quad & \quad k_{\text{D}+} &= \frac{(1-\rho) \tau_{\text{tot}} - J_c}{2\pi_0}
\end{align}
This explicit reconstruction illuminates the $\boldsymbol{\pi}$-degeneracy in the inverse design discussed in the main text. \\

\textbf{Cycle Symmetry and the Sum Theorem:} We start from the universal Cycle Symmetry (Eq.~\ref{eq:SI_Cycle_Symmetry}) evaluated for the cycle flux $\bar{x}_\alpha = J_c$:
\begin{equation}
    \sum_{(ij)\in c^{+}}\frac{1}{\pi_{i}}\frac{\partial J_c}{\partial k_{ij}} - \sum_{(ji)\in c^{-}}\frac{1}{\pi_{j}}\frac{\partial J_c}{\partial k_{ji}} = 2.
\end{equation}
The kinetic barrier perturbation operator on an edge $ij$ modifies both forward and backward rates equally in logarithmic space: $\partial_{B_{ij}} \equiv - k_{ij} \partial_{k_{ij}} - k_{ji} \partial_{k_{ji}}$. By utilizing the Edge Reciprocity symmetry (Eq.~\ref{eq:SI_Edge_Reciprocity}), $\frac{1}{\pi_i}\partial_{k_{ij}}J_c = -\frac{1}{\pi_j}\partial_{k_{ji}}J_c$, the terms in the cycle symmetry summation simplify, and the barrier operator maps to the rate sensitivity via the local net flux $J_{ij}$:
\begin{equation}
    \frac{\partial J_c}{\partial B_{ij}} = - \frac{J_{ij}}{\pi_i} \frac{\partial J_c}{\partial k_{ij}}.
\end{equation}
Substituting this mapping back into the Cycle Symmetry establishes a generic sum rule over kinetic barriers for any directed cycle:
\begin{equation}
    \sum_{(ij) \in c} \frac{J_c}{J_{ij}} \frac{\partial \ln J_c}{\partial B_{ij}} = -1.
\end{equation}
This analytical relation provides a universal constraint linking local barrier perturbations to macroscopic cycle fluxes.\\

For a unicircular process like the F$_1$-ATPase model, probability conservation ensures the net flux along every edge is identical to the overall cycle flux ($J_{\text{T}} = J_{\text{D}} = J_c$). Consequently, the coefficients $J_c/J_{ij}$ strictly evaluate to 1, collapsing the relation into a direct summation over the cycle's barriers:
\begin{equation}
    \frac{\partial \ln J_c}{\partial B_{\text{T}}} + \frac{\partial \ln J_c}{\partial B_{\text{D}}} = -1.
\end{equation}
This uniform summation is readily verified by reading the analytical sensitivities directly from the cycle flux row (row 4) of the inverse Jacobian matrix $\mathbf{A}^{-1}$ derived above:
\begin{equation}
    \left( - \frac{\Sigma_{\text{D}}}{\Sigma} \right) + \left( - \frac{\Sigma_{\text{T}}}{\Sigma} \right) = - \frac{\Sigma_{\text{D}} + \Sigma_{\text{T}}}{\Sigma} = -1.
\end{equation}
This explicitly validates the analytical inversion of the Jacobian and confirms its adherence to the network's topological constraints.\\

While this uniform summation $\sum \partial_{B} \ln J_c = -1$ serves as the stochastic network analog to the Flux Control Summation Theorem in Metabolic Control Analysis, it holds exclusively for unicircular processes. In multicyclic networks, the localized net fluxes $J_{ij}$ diverge from the target cycle flux $J_c$, introducing edge-specific weighting coefficients ($J_c/J_{ij}$) into the barrier sum rule. Therefore, the foundational Cycle Symmetry constraint (Eq.~\ref{eq:SI_Cycle_Symmetry}) remains the more general and fundamental topological law.

\subsection{Response Relations and Schur Complement Inversion}

To map the topological routing responses of the F$_1$-ATPase, we evaluate its Jacobian matrix $\mathbf{A}$ and its inverse $\mathbf{A}^{-1}$ \cite{yang_fluctuation-response_2026}. Recall from Section I.B that the Jacobian matrix is formally defined as $\mathbf{A}_{(ij),\alpha} \equiv k_{ij} \frac{\partial \mathfrak{F}_\alpha}{\partial k_{ij}}$, which links the microscopic rate perturbations to the macroscopic force responses. By choosing the rate representations $\mathbf{k}=(k_{\text{T}+},k_{\text{D}-},k_{\text{T}-},k_{\text{D}+})$ and their corresponding conjugate forces $\boldsymbol{\mathfrak{F}}=(\mathfrak{F}_\text{edge,T},\mathfrak{F}_\text{edge,D},\mathfrak{F}_\text{node,1},\mathfrak{F}_\text{cycle})$, the analytical Jacobian matrix evaluates to:
\begin{equation}
    \mathbf{A}=
    \begin{bmatrix}
    1/2 & 0 & k_{\text{T}+} & 1/2\\
    0 & 1/2 & -k_{\text{D}-} & 1/2\\
    1/2 & 0 & -k_{\text{T}-} & -1/2\\
    0 & 1/2 & k_{\text{D}+} & -1/2
    \end{bmatrix}.
\end{equation}
Because of the network's local-global topological constraints, this sparse matrix possesses a highly symmetric block structure.\\

To invert this analytically, we partition the matrix into four $2 \times 2$ blocks: $\mathbf{A} = \begin{bmatrix} \mathbf{P} & \mathbf{Q} \\ \mathbf{R} & \mathbf{S} \end{bmatrix}$. Because the upper-left block $\mathbf{P}$ and the lower-left block $\mathbf{R}$ are simply diagonal matrices ($\mathbf{P} = \mathbf{R} = \frac{1}{2}\mathbf{I}$), their products simplify immensely: $\mathbf{P}^{-1} = 2\mathbf{I}$ and $\mathbf{R}\mathbf{P}^{-1} = \mathbf{I}$. Using the Schur complement $\mathbf{M} = \mathbf{S} - \mathbf{R}\mathbf{P}^{-1}\mathbf{Q} = \mathbf{S} - \mathbf{Q}$, we find:
\begin{equation}
    \mathbf{M} = 
    \begin{bmatrix}
    -k_{\text{T}-} & -1/2\\
    k_{\text{D}+} & -1/2
    \end{bmatrix} - 
    \begin{bmatrix}
    k_{\text{T}+} & 1/2\\
    -k_{\text{D}-} & 1/2
    \end{bmatrix} = 
    \begin{bmatrix}
    -\Sigma_{\text{T}} & -1 \\
    \Sigma_{\text{D}} & -1
    \end{bmatrix},
\end{equation}
where $\Sigma_{\text{X}} = k_{\text{X}+} + k_{\text{X}-}$ denotes the sum of rates on edge $\text{X}$. The determinant of the Schur complement is exactly the sum of all rates: $\det(\mathbf{M}) = \Sigma_{\text{T}} + \Sigma_{\text{D}} = \Sigma_{\text{rates}}$. \\

Inverting $\mathbf{M}$ and applying the standard block matrix inversion formula yields the exact analytical inverse:
\begin{align}
\mathbf{A}^{-1} &= \begin{bmatrix} \mathbf{P}^{-1} + \mathbf{P}^{-1}\mathbf{Q}\mathbf{M}^{-1} & -\mathbf{P}^{-1}\mathbf{Q}\mathbf{M}^{-1} \\ -\mathbf{M}^{-1} & \mathbf{M}^{-1} \end{bmatrix}\\
     &= \frac{1}{\Sigma_{\text{rates}}}
    \begin{bmatrix}
    \Sigma_{\text{D}}+2k_{\text{T}-} & \Delta_{\text{T}} & \Sigma_{\text{D}}+2k_{\text{T}+} & -\Delta_{\text{T}}\\
    -\Delta_{\text{D}} & \Sigma_{\text{T}}+2k_{\text{D}+} & \Delta_{\text{D}} & \Sigma_{\text{T}}+2k_{\text{D}-}\\
    1 & -1 & -1 & 1\\
    \Sigma_{\text{D}} & \Sigma_{\text{T}} & -\Sigma_{\text{D}} & -\Sigma_{\text{T}}
    \end{bmatrix},
\end{align}
where $\Delta_{\text{X}} = k_{\text{X}+} - k_{\text{X}-}$. For the average observables $\bar{\textbf{x}}=(\tau_{\text{T}},\tau_{\text{D}},\pi_1,J_c)$, this inverse matrix directly yields the exact rate sensitivities via the RIM relation (Eq.~\ref{eq:SI_RIM}), $\partial_{k_{ij}} \bar{x}_\alpha = \pi_i [\mathbf{A}^{-1}]_{\alpha,(ij)}$, providing the explicit basis for the barrier sensitivity results discussed in the main text.

\section{Example 2: Kinetic Proofreading and the Performance Envelope}

\subsection{Construction of mixed coordinates and rate mapping}

To evaluate the operational space of kinetic proofreading (KP) networks, we map the functional performance metrics to the transition rates using the Caliber Force Theory (CFT) mixed coordinate system. We model both the T7 DNA polymerase and the \textit{E. coli} ribosome using a 5-state network: the empty enzyme ($\text{E}$), the initial cognate/non-cognate bound states ($\text{ER}, \text{EW}$), and the activated/catalytic states ($\text{ER}^*, \text{EW}^*$).\\

\begin{figure}
    \centering
    \includegraphics[width=\linewidth]{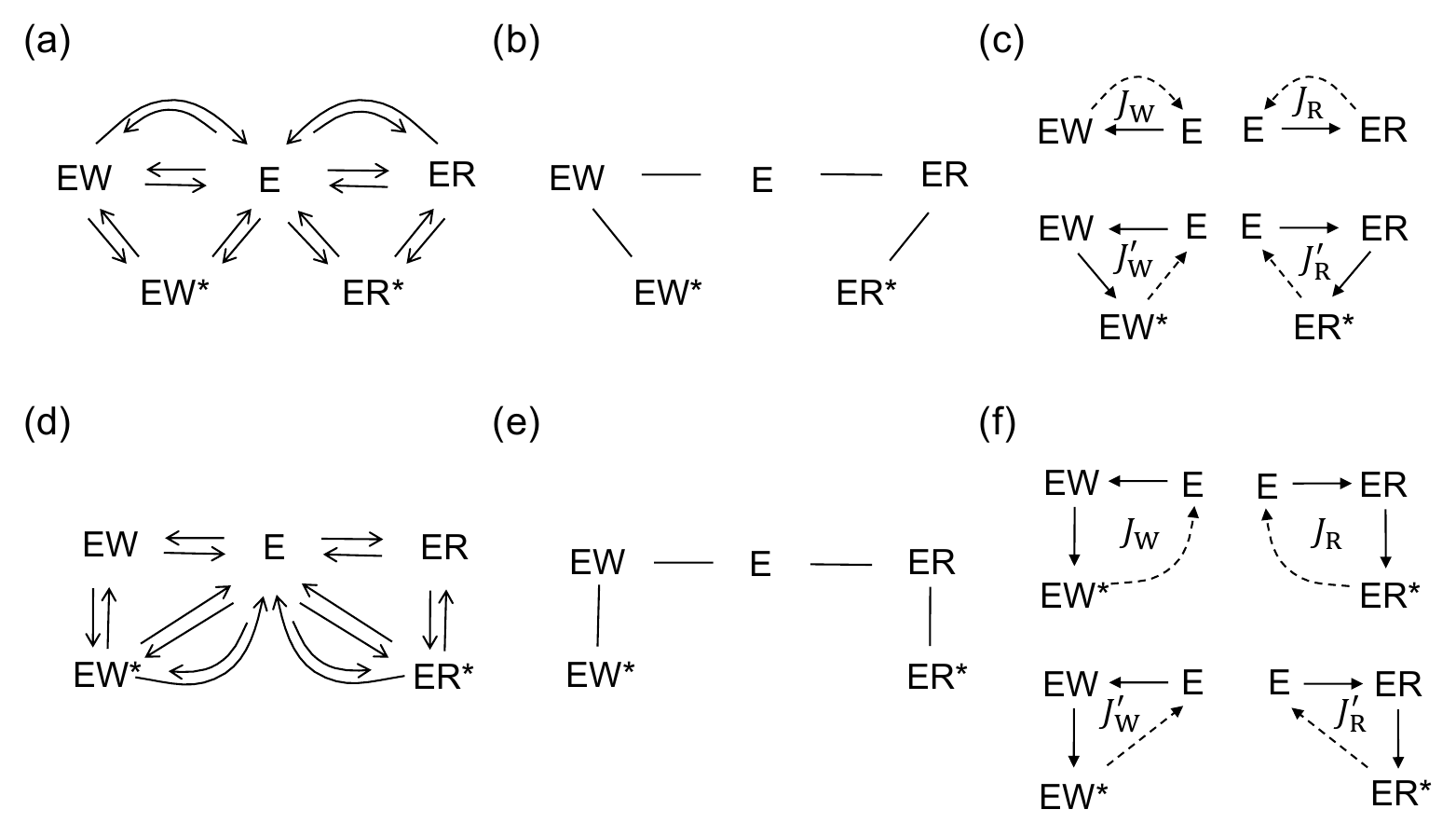}
    \caption{\textbf{Network topologies, spanning trees, and fundamental cycles for the kinetic proofreading models.} (a) The 5-state network state space for the T7 DNA polymerase. (b) The chosen spanning tree (solid lines) consisting of the backbone transitions. (c) The fundamental cycles formed by adding the chord transitions (dashed arrows) and removing the redundant edges in the spanning tree. These chords define the cognate/non-cognate incorporation ($J_\text{R}$, $J_\text{W}$) and proofreading discard ($J'_\text{R}$, $J'_\text{W}$) fluxes. (d-f) The corresponding network state space, spanning tree, and fundamental cycles for the \textit{E. coli} ribosome, highlighting the distinct topological origin of its productive incorporation transitions.}
    \label{fig: cycles and trees}
\end{figure}

As illustrated in Fig.~\ref{fig: cycles and trees}, to construct a linearly independent basis for net fluxes, we select a spanning tree consisting of the backbone transitions: $\text{E} \rightleftharpoons \text{ER} \rightleftharpoons \text{ER}^*$ and $\text{E} \rightleftharpoons \text{EW} \rightleftharpoons \text{EW}^*$. The four chords completing the fundamental cycles correspond to the irreversible product release and proofreading discard transitions. Based on the distinct topologies of the two machines, the fundamental cycle fluxes are identified as:
\begin{enumerate}
    \item $J_\text{R}$: Cognate incorporation flux (chord $\text{ER} \to \text{E}$ for polymerase; $\text{ER}^* \to \text{E}$ for ribosome).
    \item $J'_\text{R}$: Cognate proofreading discard flux (chord $\text{ER}^* \to \text{E}$ for both machines).
    \item $J_\text{W}$: Non-cognate incorporation flux (chord $\text{EW} \to \text{E}$ for polymerase; $\text{EW}^* \to \text{E}$ for ribosome).
    \item $J'_\text{W}$: Non-cognate proofreading discard flux (chord $\text{EW}^* \to \text{E}$ for both machines).
\end{enumerate}
The performance metrics are algebraically defined by these fundamental fluxes:
\begin{align}
    S &= J_\text{R} + J_\text{W}, \label{eq:SI_KP_S}\\
    \varepsilon &= \log_{10}\left(\frac{J_\text{W}}{J_\text{R}}\right), \label{eq:SI_KP_eps}\\
    C &= S \Delta\mu_{\text{p}} + (J'_\text{R} + J'_\text{W}) \Delta\mu_{\text{NTP}}, \label{eq:SI_KP_C}\\
    r' &= \frac{J'_\text{W}}{J'_\text{R}}. \label{eq:SI_KP_r}
\end{align}
Here, $\Delta\mu_{\text{p}}$ is the free energy of bond formation and $\Delta\mu_{\text{NTP}}$ is the free energy of NTP hydrolysis. Assuming the network operates in a functional regime where both incorporation fluxes are positive ($J_\text{R} > 0, J_\text{W} > 0$), we invert this system to explicitly solve for the fundamental cycle fluxes as functions of the target coordinate $(S, \varepsilon, C, r')$:
\begin{align}
    J_\text{R} &= \frac{S}{1 + 10^\varepsilon}, \quad J_\text{W} = \frac{S \cdot 10^\varepsilon}{1 + 10^\varepsilon}, \label{eq:SI_KP_J_incorp}\\
    J'_\text{R} &= \frac{C - S\Delta\mu_{\text{p}}}{\Delta\mu_{\text{NTP}}(1 + r')}, \quad J'_\text{W} = \frac{r'(C - S\Delta\mu_{\text{p}})}{\Delta\mu_{\text{NTP}}(1 + r')}. \label{eq:SI_KP_J_discard}
\end{align}
These explicit relations are crucial for substituting the abstract cycle fluxes with tangible functional targets in our inverse design coordinate.\\

\textbf{Coordinate Swap via Cycle Forces:} The key to enabling inverse design is transforming the ``microcanonical'' coordinate $\bar{\textbf{x}} = (\boldsymbol{\pi}, \boldsymbol{\tau}, \boldsymbol{J})$ into a canonical-like coordinate where the cycle forces $\boldsymbol{\mathfrak{F}}_{\text{cycle}}$ are fixed. For any fundamental cycle $c$, the cycle force is the sum of its edge affinities:
\begin{equation}
    \mathfrak{F}_{\text{cycle},c} = \text{arctanh}\left(\frac{J_c}{\tau_c}\right) + \sum_{(ij) \in c \setminus \{c\}} \text{arctanh}\left(\frac{J_{ij}}{\tau_{ij}}\right),
\end{equation}
where $\tau_c$ is the traffic exclusively on the defining chord. Because the $\text{arctanh}$ function is strictly monotonic and invertible, we can analytically isolate $\tau_c$:
\begin{equation} \label{eq:SI_KP_Chord_Traffic}
    \tau_c = J_c \coth\left( \mathfrak{F}_{\text{cycle},c} - \sum_{(ij) \in c \setminus \{c\}} \text{arctanh}\left(\frac{J_{ij}}{\tau_{ij}}\right) \right).
\end{equation}
This explicit substitution removes the four chord traffics from the set of independent variables. The remaining edge traffics along the spanning tree are denoted as $\boldsymbol{\tau}_{\text{rem}}$. This establishes the exact analytical mapping into the mixed coordinate system $(\boldsymbol{\pi}, \boldsymbol{\tau}_{\text{rem}}, \boldsymbol{J}, \boldsymbol{\mathfrak{F}}_{\text{cycle}})$. \\

Furthermore, because the four cycle forces are explicitly constrained by the thermodynamic driving forces ($2\mathfrak{F}_{\text{cycle, R}} = 2\mathfrak{F}_{\text{cycle, W}} = \Delta\mu_{\text{p}}$ for the incorporation cycles, and $2\mathfrak{F}'_{\text{cycle, R}} = 2\mathfrak{F}'_{\text{cycle, W}} = \Delta\mu_{\text{NTP}}$ for the discard cycles), and the fundamental cycle fluxes $\boldsymbol{J}$ can be fully parameterized by the performance targets $(S, \varepsilon, C, r')$ via Eqs.~\eqref{eq:SI_KP_J_incorp}--\eqref{eq:SI_KP_J_discard}, we ultimately construct the inverse design coordinate: $(\boldsymbol{\pi}, \boldsymbol{\tau}_{\text{rem}}, S, \varepsilon, C, r', \boldsymbol{\mathfrak{F}}_{\text{cycle}})$. \\

\textbf{Systematic Rate Reconstruction:} To translate an operational target point in this functional coordinate space back to the 16 microscopic transition rates, we execute the following sequence:
\begin{enumerate}
    \item Evaluate the four fundamental cycle fluxes $J_c \in \{J_\text{R}, J_\text{W}, J'_\text{R}, J'_\text{W}\}$ from the target metrics $(S, \varepsilon, C, r')$ using Eqs.~\eqref{eq:SI_KP_J_incorp} and \eqref{eq:SI_KP_J_discard}.
    \item Compute the local net flux $J_{ij}$ on every edge by superimposing the fundamental cycle fluxes traversing it. For instance, the initial cognate binding edge $\text{E} \rightleftharpoons \text{ER}$ participates in both the incorporation and discard cycles, yielding $J_{\text{E} \to \text{ER}} = J_\text{R} + J'_\text{R}$. As another example mapped in Fig.~\ref{fig: cycles and trees}, the activation edge $\text{ER} \rightleftharpoons \text{ER}^*$ in the T7 DNA polymerase carries exclusively the proofreading discard flux ($J_{\text{ER} \to \text{ER}^*} = J'_\text{R}$), whereas the same edge in the \textit{E. coli} ribosome carries both fluxes ($J_{\text{ER} \to \text{ER}^*} = J_\text{R} + J'_\text{R}$).
    \item With all local fluxes $J_{ij}$ determined, substitute the fixed cycle forces $\boldsymbol{\mathfrak{F}}_{\text{cycle}}$ and the remaining tree traffics $\boldsymbol{\tau}_{\text{rem}}$ into Eq.~\eqref{eq:SI_KP_Chord_Traffic} to algebraically resolve the four dependent chord traffics $\tau_c$.
    \item Having completely specified all state probabilities $\pi_i$, edge traffics $\tau_{ij}$, and local fluxes $J_{ij}$, the 16 exact microscopic transition rates are reconstructed via the fundamental CFT mapping:
    \begin{equation} \label{eq:SI_KP_rate_reconstruction}
        k_{ij} = \frac{\tau_{ij} + J_{ij}}{2\pi_i}, \quad k_{ji} = \frac{\tau_{ij} - J_{ij}}{2\pi_j}.
    \end{equation}
\end{enumerate}
This sequence perfectly isolates the physical parameters, allowing us to explicitly design the microscopic transition rates that realize any mathematically feasible set of performance targets.

\subsection{Traffic bounds and anti-proofreading regimes}

The rate reconstruction in Eq.~\eqref{eq:SI_KP_rate_reconstruction} is physically bounded by the non-negativity of the one-way probability fluxes, $p_{ij} = (\tau_{ij} + J_{ij})/2 \ge 0$. This imposes the strict physical constraint $|J_{ij}| \le \tau_{ij}$ on every edge.\\ 

Projecting the limit $J_{\text{E} \to \text{ER}} \le \tau_{\text{E, ER}}$ into the functional coordinate space using Eqs.~\eqref{eq:SI_KP_J_incorp} and \eqref{eq:SI_KP_J_discard} yields the trade-off bound presented in the main text:
\begin{equation} \label{eq:SI_KP_bound_R}
    \underbrace{\frac{S}{1 + 10^\varepsilon}}_{J_R} + \underbrace{\frac{C - S\Delta\mu_{\text{p}}}{\Delta\mu_{\text{NTP}}(1 + r')}}_{J'_R} \le \tau_{\text{E, ER}}.
\end{equation}
This inequality demonstrates a kinetic traffic jam: the productive incorporation speed and the proofreading dissipation must compete for the shared finite bandwidth $\tau_{\text{E, ER}}$. Similarly, the flux limit on the non-cognate binding edge, $J_{\text{E} \to \text{EW}} \le \tau_{\text{E, EW}}$, establishes a parallel boundary for the non-cognate pathway:
\begin{equation} \label{eq:SI_KP_bound_W}
    \underbrace{\frac{S \cdot 10^\varepsilon}{1 + 10^\varepsilon}}_{J_\text{W}} + \underbrace{\frac{r'(C - S\Delta\mu_{\text{p}})}{\Delta\mu_{\text{NTP}}(1 + r')}}_{J'_\text{W}} \le \tau_{\text{E, EW}}.
\end{equation}
Furthermore, the independent chord transitions impose strict boundaries on the individual discard fluxes:$|J'_\text{R}| \le \tau_{\text{ER}^*, \text{E}}, \quad |J'_\text{W}| \le \tau_{\text{EW}^*, \text{E}}$.\\

\textbf{Anti-Proofreading Regime:} The standard proofreading regime operates under positive NTP dissipation ($C \ge S\Delta\mu_{\text{p}}$), yielding $J'_\text{R}, J'_\text{W} \ge 0$. However, driving the energetic cost $C$ below the incorporation baseline $S\Delta\mu_{\text{p}}$ forces the net discard fluxes to become negative. In this anti-proofreading regime, the network must run the discard cycles in reverse to synthesize NTP. \\

While Eq.~\eqref{eq:SI_KP_bound_W} represents the upper capacity bound, a highly negative non-cognate discard flux triggers the corresponding lower bound constraint:
\begin{equation}
    J_\text{W} + J'_\text{W} \ge -\tau_{\text{E, EW}}.
\end{equation}
When the reversed non-cognate discard flux strictly exceeds the available binding traffic bandwidth ($J'_\text{W} < -\tau_{\text{E, EW}}$, or equivalently $|J'_\text{W}| > \tau_{\text{E, EW}}$), we substitute $J_\text{W}$ and $-|J'_\text{W}|$ into the lower bound:
\begin{equation}
    S \frac{10^\varepsilon}{1 + 10^\varepsilon} - |J'_\text{W}| \ge -\tau_{\text{E, EW}}.
\end{equation}
Multiplying both sides by $(1 + 10^\varepsilon)$ and isolating the $10^\varepsilon$ terms yields the intermediate step:
\begin{equation}
    10^\varepsilon \left( S - |J'_\text{W}| + \tau_{\text{E, EW}} \right) \ge |J'_\text{W}| - \tau_{\text{E, EW}}.
\end{equation}
By substituting $-|J'_\text{W}| = J'_\text{W}$ back into the left-hand parenthesis, this algebraically reorganizes into a strict lower bound on the error rate:
\begin{equation}
    10^\varepsilon \ge \frac{|J'_\text{W}| - \tau_{\text{E, EW}}}{S + J'_\text{W} + \tau_{\text{E, EW}}}.
\end{equation}
This establishes a minimum achievable error floor, revealing a topological constraint when the system is driven into regimes of negative proofreading dissipation.

\subsection{Algorithmic details of the linear exploration}

To navigate the functional interior without violating these physical limits, we inversely designed an optimization protocol (Main Text Fig.~5c, d) operating directly within our key coordinate $(\boldsymbol{\pi}, \boldsymbol{\tau}_{\text{rem}}, S, \varepsilon, C, r', \boldsymbol{\mathfrak{F}}_{\text{cycle}})$. The algorithm explores the concurrent improvement of speed, accuracy, and cost as follows:

\begin{enumerate}
    \item \textbf{Initialization:} The coordinate is initialized at the wild-type parameters $\mathbf{z}_{\text{WT}}$ derived from Mallory \textit{et al.} \cite{mallory_trade-offs_2019}, explicitly listed in Table~\ref{tab:SI_WT_Params}. The state probabilities $\boldsymbol{\pi}$, the spanning-tree traffics $\boldsymbol{\tau}_{\text{rem}}$, and the discard ratio $r'$ are clamped to their wild-type values.
    \item \textbf{Linear Progression:} The target functional variables are driven along a linear sequence parameterized by $\lambda$:
    \begin{equation}
        S(\lambda) = S_{\text{WT}} + \lambda \Delta S, \quad \varepsilon(\lambda) = \varepsilon_{\text{WT}} - \lambda \Delta\varepsilon, \quad C(\lambda) = C_{\text{WT}} - \lambda \Delta C.
    \end{equation}
    For the explicit optimization trajectories demonstrated in the main text, the directional vector is set to $\Delta S = 3 \times 10^{-7}\ \text{s}^{-1}$, $\Delta \varepsilon = -0.02$, and $\Delta C = -1.0\ k_{\text{B}}T$.
    \item \textbf{Traffic Relaxation Rule:} At each step $\lambda$, the required local fluxes $J_{ij}(\lambda)$ are evaluated. We continuously monitor the saturation ratio $|J_{ij}| / \tau_{ij}$ for all edges. If any edge collides with the physical boundary ($|J_{ij}| / \tau_{ij} \ge 1.0$), the local kinetic bandwidth is expanded by multiplying the traffic of that specific edge by 10 ($\tau_{ij} \to 10 \tau_{ij}$). This allows the free-lunch optimization trajectory to proceed beyond the initial wild-type kinetic envelope.
    \item \textbf{Rate Execution:} With the dynamically updated traffics and fluxes, the transition rates $k_{ij}(\lambda)$ are generated via the analytical mapping in Eq.~\eqref{eq:SI_KP_rate_reconstruction}. 
\end{enumerate}
The abrupt jumps observed in the engineered transition rates (Main Text Fig.~5d) are the direct consequence of these traffic relaxation events triggering to prevent negative probability fluxes.

\begin{table}
\centering
\begin{tabular}{l c c | c c}
\hline \hline
 & \multicolumn{2}{c|}{\textbf{T7 DNA Polymerase}} & \multicolumn{2}{c}{\textbf{\textit{E. coli} Ribosome}} \\
\textbf{Transition (s$^{-1}$)} & \textbf{Cognate ($k_{\text{R}}$)} & \textbf{Non-Cognate ($k_{\text{W}}$)} & \textbf{Cognate ($k_{\text{R}}$)} & \textbf{Non-Cognate ($k_{\text{W}}$)} \\ \hline
$k_1$ ($\text{E} \to \text{EX}$) & $250.0$ & $2.0 \times 10^{-3}$ & $40.0$ & $27.0$ \\
$k_{-1}$ ($\text{EX} \to \text{E}$) & $1.0$ & $1.0 \times 10^{-5}$ & $0.5$ & $47.0$ \\
$k_2$ ($\text{EX} \to \text{EX}^*$) & $0.2$ & $2.3$ & $25.0$ & $1.2$ \\
$k_{-2}$ ($\text{EX}^* \to \text{EX}$) & $700.0$ & $700.0$ & $1.0 \times 10^{-3}$ & $1.0 \times 10^{-3}$ \\
$k_3$ (Discard) & $900.0$ & $900.0$ & $8.5 \times 10^{-2}$ & $0.6715$ \\
$k_{-3}$ (Reverse Discard) & $1.32 \times 10^{-7}$ * & $1.22 \times 10^{-9}$ * & $3.50 \times 10^{-4}$ * & $9.54 \times 10^{-7}$ * \\
$k_p$ (Incorporation) & $250.0$ & $1.2 \times 10^{-2}$ & $8.415$ & $3.53 \times 10^{-2}$ \\
$k_{-p}$ (Reverse Incorp.) & $3.20 \times 10^{-7}$ * & $1.23 \times 10^{-11}$ * & $8.60 \times 10^{-5}$ * & $1.24 \times 10^{-10}$ * \\
\hline \hline
\end{tabular}
\caption{\textbf{Explicit wild-type transition rates for the kinetic proofreading models.} All 16 microscopic transition rates for the cognate (Right) and non-cognate (Wrong) pathways are explicitly listed to ensure theoretical reproducibility, adapted from Mallory \textit{et al.} \cite{mallory_trade-offs_2019}. Rates marked with an asterisk (*) were computed to enforce local detailed balance. Specifically, the reverse proofreading discard rates were analytically constrained by the free energy of NTP hydrolysis ($\Delta\mu_{\text{NTP}} = 20.0~k_{\text{B}}T$) via $k_{-3} = (k_1 k_2 k_3)/(k_{-1} k_{-2} e^{\Delta\mu_{\text{NTP}}})$. Similarly, the reverse incorporation rates were constrained by the bond formation free energy ($\Delta\mu_{\text{p}} = 26.0~k_{\text{B}}T$). Due to the distinct topologies of the production cycles, $k_{-p} = (k_1 k_p)/(k_{-1} e^{\Delta\mu_{\text{p}}})$ for the T7 DNA polymerase, whereas $k_{-p} = (k_1 k_2 k_p)/(k_{-1} k_{-2} e^{\Delta\mu_{\text{p}}})$ for the \textit{E. coli} ribosome.}
\label{tab:SI_WT_Params}
\end{table}

\section{Example 3: Enzyme Inhibition and Non-Invertible Networks}

\subsection{Caliber Force construction for non-invertible networks}

In this section, we provide the construction of the Caliber Force Theory (CFT) framework for the noncompetitive inhibition model. This model is unique because it incorporates a non-invertible catalytic step, representing the chemical production of the enzyme. The Maximum Caliber framework maximizes the path entropy relative to a reference process. The resulting parameterization of path probability is still an exponential family, and the conjugate coordinates still exist and are still converted by Legendre-Fenchel transforms.\\

\textbf{1. Model Topology and Transition Matrix:} We consider a four-state enzyme system: free enzyme $\text{E}$ (node 0), enzyme-substrate complex $\text{ES}$ (node 1), enzyme-inhibitor-substrate complex $\text{EIS}$ (node 2), and enzyme-inhibitor complex $\text{EI}$ (node 3). The network is connected by invertible binding transitions, but the production step is non-invertible: $\text{ES} \xrightarrow{k_{\text{cat}}} \text{E}$ ($1 \to 0$).
The transition rate matrix $\mathbf{k}$ describes the rate of probability flow between these states:
\begin{equation}
\mathbf{k} = \begin{bmatrix}
-\sum_{i}k_{0i} & k_{01} & 0 & k_{03}\\
k_{10}+k_{\text{cat}} & -\sum_{j}k_{1j}-k_{\text{cat}} & k_{12} & 0\\
0 & k_{21} & -\sum_{l}k_{2l} & k_{23}\\
k_{30} & 0 & k_{32} & -\sum_{m}k_{3m}
\end{bmatrix}.
\end{equation}
This explicit matrix formulation serves as the mathematical basis for extracting the conjugate forces.\\

\textbf{2. Observable Basis and Fundamental Cycles:} To construct the linearly independent observable basis $\bar{\textbf{x}}$, we select node $0$ ($\text{E}$) as the reference node $m$, making $\pi_1, \pi_2, \pi_3$ the independent state probabilities. We then select a spanning tree covering the transitions $1 \rightleftharpoons 0 \rightleftharpoons 3 \rightleftharpoons 2$. The fundamental cycles are defined by adding the remaining chords. The reversible binding edge $1 \rightleftharpoons 2$ forms the invertible leak cycle ($0 \rightleftharpoons 3 \rightleftharpoons 2 \rightleftharpoons 1 \rightleftharpoons 0$), with its net flux denoted as $J_{\text{leak}} = J_{21}$. The non-invertible catalytic edge $1 \to 0$ acts as the defining chord for the production cycle ($0 \rightleftharpoons 1 \to 0$), carrying the flux $J_{\text{cat}}$. Because the catalytic step is non-invertible, its symmetric traffic $\tau_{\text{cat}}$ strictly equals its directed flux $J_{\text{cat}}$ and is thus removed from the set of independent variables. The resulting observable coordinate vector is $\bar{\textbf{x}} = (\tau_{01}, \tau_{03}, \tau_{12}, \tau_{23}, \pi_1, \pi_2, \pi_3, J_{\text{cat}}, J_{\text{leak}})$.\\

\textbf{3. Tilted Matrix and Conjugate Forces:} To identify the rate expressions of conjugate forces, we resort to the rate parameterization in Eq.~\eqref{eq: k(F)}, which requires first constructing the tilted matrix $\tilde{\mathbf{M}}$. The off-diagonal elements are tilted by the forces associated with each specific transition. The diagonal elements capture node constraints defined by the out-degree of each node in the reference ``doldrum'' process (e.g., node 1 connects to nodes 0 and 2 via reversible edges and node 0 via the catalytic edge, yielding an out-degree of 3). The largest eigenvalue of $\tilde{\mathbf{M}}$ is the Caliber $\mathfrak{c}$, which acts as the dynamical partition function:
\begin{equation}
\tilde{\mathbf{M}} = \begin{bmatrix}
-2 & e^{\mathfrak{F}_{\text{edge},01}} & 0 & e^{\mathfrak{F}_{\text{edge},03}}\\
e^{\mathfrak{F}_{\text{edge},01}} + e^{\mathfrak{F}_{\text{cat}}} & \mathfrak{F}_{\text{node},1}-3 & e^{\mathfrak{F}_{\text{edge},12}-\mathfrak{F}_{\text{leak}}} & 0\\
0 & e^{\mathfrak{F}_{\text{edge},12}+\mathfrak{F}_{\text{leak}}} & \mathfrak{F}_{\text{node},2}-2 & e^{\mathfrak{F}_{\text{edge},23}}\\
e^{\mathfrak{F}_{\text{edge},03}} & 0 & e^{\mathfrak{F}_{\text{edge},23}} & \mathfrak{F}_{\text{node},3}-2
\end{bmatrix}.
\end{equation}
The right eigenvector $\mathbf{v}$ associated with $\mathfrak{c}$ bridges the macroscopic forces back to the microscopic transition rates via $k_{ij} = \tilde{M}_{ij} v_j / v_i$. The corresponding rate matrix $\mathbf{k}$ evaluates to:
\begin{equation}
\mathbf{k} = \begin{bmatrix}
-2-\mathfrak{c} & e^{\mathfrak{F}_{\text{edge},01}}\frac{v_1}{v_0} & 0 & e^{\mathfrak{F}_{\text{edge},03}}\frac{v_3}{v_0}\\
e^{\mathfrak{F}_{\text{edge},01}}\frac{v_0}{v_1} + e^{\mathfrak{F}_{\text{cat}}}\frac{v_0}{v_1} & \mathfrak{F}_{\text{node},1}-3-\mathfrak{c} & e^{\mathfrak{F}_{\text{edge},12}-\mathfrak{F}_{\text{leak}}}\frac{v_2}{v_1} & 0\\
0 & e^{\mathfrak{F}_{\text{edge},12}+\mathfrak{F}_{\text{leak}}}\frac{v_1}{v_2} & \mathfrak{F}_{\text{node},2}-2-\mathfrak{c} & e^{\mathfrak{F}_{\text{edge},23}}\frac{v_3}{v_2}\\
e^{\mathfrak{F}_{\text{edge},03}}\frac{v_0}{v_3} & 0 & e^{\mathfrak{F}_{\text{edge},23}}\frac{v_2}{v_3} & \mathfrak{F}_{\text{node},3}-2-\mathfrak{c}
\end{bmatrix}.
\end{equation}
Evaluating the logarithmic ratios of these reconstructed rates along specific network structures completely cancels out the eigenvector components. To extract the catalytic cycle force, we evaluate $k_{\text{cat}}$ and the reversible binding rates on the $\text{E} \rightleftharpoons \text{ES}$ edge. Since $k_{\text{cat}} = e^{\mathfrak{F}_{\text{cat}}}\frac{v_0}{v_1}$ and the symmetric binding rates satisfy $\frac{k_{01}}{k_{10}} = (\frac{v_1}{v_0})^2$, taking the logarithm yields $\ln k_{\text{cat}} + \frac{1}{2}\ln\frac{k_{01}}{k_{10}} = \mathfrak{F}_{\text{cat}} + \ln\frac{v_0}{v_1} + \ln\frac{v_1}{v_0} = \mathfrak{F}_{\text{cat}}$. Repeating this for the remaining structures yields the explicit algebraic mapping from rates to conjugate forces:
\begin{align}
    \mathfrak{F}_{\text{node},n} &= \sum_{i}(k_{0i}-1) - \sum_{j}(k_{nj}-1), \\
    \mathfrak{F}_{\text{edge},ij} &= \frac{1}{2}\ln(k_{ij}k_{ji}), \\
    \mathfrak{F}_{\text{leak}} &= \frac{1}{2}\sum_{(ij) \in 0 \to 3 \to 2 \to 1 \to 0}\ln\frac{k_{ij}}{k_{ji}}, \\
    \mathfrak{F}_{\text{cat}} &= \ln k_{\text{cat}} + \frac{1}{2}\ln\frac{k_{01}}{k_{10}}.
\end{align}
This demonstrates that the inclusion of a non-invertible step only modifies the definition of the specific cycle force traversing it ($\mathfrak{F}_{\text{cat}}$), while preserving the standard force definitions for all nodes, edges, and invertible cycles ($\mathfrak{F}_{\text{leak}}$).\\

\textbf{4. The Jacobian Matrix $\mathbf{A}$:} The Jacobian matrix $\mathbf{A}_{(ij),\alpha} = k_{ij}\partial \mathfrak{F}_{\alpha} / \partial k_{ij}$ serves as the dictionary mapping local rate perturbations to global force responses. We define the rate vector $\mathbf{k} = (k_{01}, k_{10}, k_{\text{cat}}, k_{03}, k_{30}, k_{12}, k_{21}, k_{23}, k_{32})^\mathsf{T}$ as the rows, and the conjugate force vector $\boldsymbol{\mathfrak{F}} = (\mathfrak{F}_{\text{edge},01}, \mathfrak{F}_{\text{edge},03}, \mathfrak{F}_{\text{edge},12}, \mathfrak{F}_{\text{edge},23}, \mathfrak{F}_{\text{node},1}, \mathfrak{F}_{\text{node},2}, \mathfrak{F}_{\text{node},3}, \mathfrak{F}_{\text{cat}}, \mathfrak{F}_{\text{leak}})^\mathsf{T}$ as the columns. The $9 \times 9$ Jacobian is analytically evaluated as:
\begin{equation}
\mathbf{A} = \begin{bmatrix}
 1/2 & 0 & 0 & 0 & k_{01} & k_{01} & k_{01} & 1/2 & -1/2 \\
 1/2 & 0 & 0 & 0 & -k_{10} & 0 & 0 & -1/2 & 1/2 \\
 0 & 0 & 0 & 0 & -k_{\text{cat}} & 0 & 0 & 1 & 0 \\
 0 & 1/2 & 0 & 0 & k_{03} & k_{03} & k_{03} & 0 & 1/2 \\
 0 & 1/2 & 0 & 0 & 0 & 0 & -k_{30} & 0 & -1/2 \\
 0 & 0 & 1/2 & 0 & -k_{12} & 0 & 0 & 0 & -1/2 \\
 0 & 0 & 1/2 & 0 & 0 & -k_{21} & 0 & 0 & 1/2 \\
 0 & 0 & 0 & 1/2 & 0 & -k_{23} & 0 & 0 & -1/2 \\
 0 & 0 & 0 & 1/2 & 0 & 0 & -k_{32} & 0 & 1/2
\end{bmatrix}.
\end{equation}
The independent edge noise sources are defined as $\lambda_{ij} = (N_{ij} - T_i k_{ij})/t$, where $N_{ij}$ are the transition counts and $T_i$ are the dwell times. The empirical observable rates are $\mathbf{x} = (f_n, \phi_{ij}, \psi_c) = (T_n, \Phi_{ij}, \Psi_c)/t$. Because node probability normalization ($\sum_n f_n = 1$) and Kirchhoff's current law ($\sum_j N_{ij} - \sum_j N_{ji} \sim 0$) constrain the raw counts, we can algebraically substitute the counting variables $N$ and $T$ with the independent basis $\mathbf{x}$. While we omit the algebraic substitution here (detailed in Ref. \cite{yang_fluctuation-response_2026}), this demonstrates that the linear mapping between the noise sources $\boldsymbol{\lambda}$ and the fluctuating observables $\mathbf{x}$ remains governed by this Jacobian: $\boldsymbol{\lambda} \sim \mathbf{A}\mathbf{x} - \nabla_{\ln \mathbf{k}} \mathfrak{c}$. Because this stochastic mapping holds, the derivation of the Response-Inverse-Matrix (RIM) relation ($\partial \bar{x}_\alpha / \partial \ln k_{ij} = \pi_i \mathbf{A}^{-1}_{\alpha, (ij)}$) provided in Eq.~\eqref{eq:SI_RIM} remains valid for networks with non-invertible transitions.

\subsection{Explicit derivation of the Inhibitor Equal Traffic Principle}

\textbf{1. Flux Conservation and Force Balance:}
The inner loop $\text{E} \rightleftharpoons \text{ES} \rightleftharpoons \text{EIS} \rightleftharpoons \text{EI} \rightleftharpoons \text{E}$ is futile with no net effect, requiring its cycle force to vanish ($\mathfrak{F}_{\text{leak}} = 0$). However, the irreversible catalytic sink at state 1 ($\text{ES} \to \text{E}$) induces a steady-state compensatory backflow, $J_{\text{leak}}$.\\

Applying Kirchhoff's Current Law at each node dictates the flow routing:
\begin{enumerate}
    \item At node 1 ($\text{ES}$): The net flux from node 0 ($J_{01}$) must satisfy $J_{01} = J_{\text{cat}} - J_{\text{leak}}$. 
    \item At the sideway nodes (2 and 3): The net fluxes $J_{21}, J_{32}, J_{03}$ are equal to $J_{\text{leak}}$, which strictly implies $J_{12}, J_{23}, J_{30} = -J_{\text{leak}}$.
\end{enumerate}
Denoting the edge affinity $\mathcal{A}_{ij} = \text{arctanh}(J_{ij}/\tau_{ij})$, the zero cycle force constraint $\sum_{(ij) \in \text{cycle}} \mathcal{A}_{ij} = 0$ along the loop $0 \to 1 \to 2 \to 3 \to 0$ is expressed as:
\begin{equation} \label{eq:SI_Inhib_full_balance}
    \text{arctanh}\left(\frac{J_{\text{cat}} - J_{\text{leak}}}{\tau_{01}}\right) - \left[ \text{arctanh}\left(\frac{J_{\text{leak}}}{\tau_{12}}\right) + \text{arctanh}\left(\frac{J_{\text{leak}}}{\tau_{23}}\right) + \text{arctanh}\left(\frac{J_{\text{leak}}}{\tau_{30}}\right) \right] = 0.
\end{equation}
Let $\mathcal{A}_{\text{side}} = \sum_{k \in \{12, 23, 30\}} \text{arctanh}(J_{\text{leak}}/\tau_k)$. Applying the hyperbolic tangent to both sides of Eq.~\eqref{eq:SI_Inhib_full_balance} isolates the catalytic production:
\begin{equation} \label{eq:SI_Inhib_derivation_step}
    \frac{J_{\text{cat}} - J_{\text{leak}}}{\tau_{01}} = \tanh(\mathcal{A}_{\text{side}}) \implies J_{\text{cat}} = J_{\text{leak}} + \tau_{01} \tanh(\mathcal{A}_{\text{side}}).
\end{equation}
This algebraic isolation of $J_{\text{cat}}$ establishes the coordinate transformation. We start from the observable coordinate $(\boldsymbol{\pi}, \boldsymbol{\tau}, J_{\text{cat}}, J_{\text{leak}})$. By enforcing the thermodynamic constraint $\mathfrak{F}_{\text{leak}} = 0$, we substitute $J_{\text{cat}}$ out of the independent variables, yielding the mixed coordinate $\mathbf{z} = (\boldsymbol{\pi}, \boldsymbol{\tau}, J_{\text{leak}}, \mathfrak{F}_{\text{leak}})$. As demonstrated in Example 1, deriving this explicit analytical solution for a flux under a fixed force constraint is equivalent to computing the derivative of the corresponding Legendre-Fenchel dynamic free energy potential.\\

\textbf{2. The Positivity of $\boldsymbol{J_\text{leak}}$:} 
The requirement that the compensatory leak flux is strictly positive ($J_{\text{leak}} > 0$) is a direct consequence of probability conservation combined with the thermodynamic constraint of the futile loop ($\mathfrak{F}_{\text{leak}} = 0$). Because the catalytic step is strictly non-invertible ($k_{\text{cat}} > 0$), it acts as a constant sink, driving a unidirectional probability flux $p_{\text{cat}} = \pi_1 k_{\text{cat}} > 0$ from node 1 ($\text{ES}$) to node 0 ($\text{E}$). To maintain a steady state, node 1 must continuously receive a net incoming probability flux from its connected pathways to replenish this loss. Kirchhoff's current law thus requires $J_{01} + J_{21} = p_{\text{cat}} > 0$. Since the unbranched sideway route enforces $J_{21} = J_{\text{leak}}$, we obtain the flux conservation condition:
\begin{equation}
    J_{01} + J_{\text{leak}} > 0.
\end{equation}
This physical boundary dictates that the sum of the primary and leak fluxes must be strictly positive.\\

Furthermore, the flows $J_{01}$ and $J_{\text{leak}}$ are bound by the zero cycle force constraint along the inhibitor loop. As derived in Eq.~\eqref{eq:SI_Inhib_full_balance}, the affinity of the primary edge must perfectly balance the total affinity of the sideway edges:
\begin{equation}
    \text{arctanh}\left(\frac{J_{01}}{\tau_{01}}\right) = \text{arctanh}\left(\frac{J_{\text{leak}}}{\tau_{12}}\right) + \text{arctanh}\left(\frac{J_{\text{leak}}}{\tau_{23}}\right) + \text{arctanh}\left(\frac{J_{\text{leak}}}{\tau_{30}}\right).
\end{equation}
The function $\text{arctanh}(x)$ is strictly monotonically increasing and odd, which means it preserves the sign of its argument. Therefore, the total sum on the right-hand side shares the exact same sign as $J_{\text{leak}}$. For the thermodynamic equality to hold, $J_{01}$ and $J_{\text{leak}}$ must strictly share the same sign (or both be zero).\\

Since $J_{01}$ and $J_{\text{leak}}$ must have the same sign, and their sum must be strictly positive to replenish the catalytic sink ($J_{01} + J_{\text{leak}} > 0$), they cannot be zero or negative. Both fluxes must therefore be strictly positive. This guarantees that the noncompetitive inhibitor loop inevitably acts as a forward compensatory shunt ($J_{\text{leak}} > 0$).\\

\textbf{3. Proof of Monotonic Coupling via Chain Rule:}
We evaluate the sensitivity of production to the leak, $\partial J_{\text{cat}} / \partial J_{\text{leak}}$, within the mixed coordinate $\mathbf{z} = (\boldsymbol{\pi}, \boldsymbol{\tau}, J_{\text{leak}}, \mathfrak{F}_{\text{leak}})$. This dictates that the partial derivative is taken while holding the kinetic traffic $\boldsymbol{\tau}$ and the cycle force $\mathfrak{F}_{\text{leak}} = 0$ constant. Using the derivative rules $\frac{d}{dx} \tanh(u) = \text{sech}^2(u) \frac{du}{dx}$ and $\frac{d}{dx} \text{arctanh}(x/a) = \frac{a}{a^2 - x^2}$, the derivative of the sideway affinity evaluates to:
\begin{equation}
    \frac{\partial \mathcal{A}_{\text{side}}}{\partial J_{\text{leak}}} = \sum_{k \in \text{side}} \frac{\tau_k}{\tau_k^2 - J_{\text{leak}}^2}.
\end{equation}
Thus, the total derivative of the catalytic flux is:
\begin{equation}
    \frac{\partial J_{\text{cat}}}{\partial J_{\text{leak}}} = 1 + \tau_{01} \underbrace{\text{sech}^2(\mathcal{A}_{\text{side}})}_{>0} \underbrace{\left( \sum_{k \in \text{side}} \frac{\tau_k}{\tau_k^2 - J_{\text{leak}}^2} \right)}_{>0}.
\end{equation}
Since all edge traffics bound their net fluxes ($\tau_k > |J_{\text{leak}}|$) for any physical NESS, the derivative $\partial J_{\text{cat}} / \partial J_{\text{leak}}$ is greater than 1. This proves that $J_{\text{cat}}$ and $J_{\text{leak}}$ are locked in a positive coupling.\\

\textbf{4. Jensen's Inequality and Global Bounds on Production:}
To identify the limits of inhibitor efficacy, we analyze how the distribution of sideway traffic $\tau_k$ modulates the production flux $J_{\text{cat}}$ under a fixed total traffic budget $\sum \tau_k = 3\bar{\tau}$.\\

The resistance function $f(\tau) = \text{arctanh}(J_{\text{leak}}/\tau)$ is convex ($f''(\tau) > 0$) for all physical regimes where $\tau > J_{\text{leak}}$. This convexity leads to two bounds:\\

\textit{(A) The Lower Bound ($J_{\text{cat}}^{\text{min}}$) --- The Inhibitor Equal Traffic Principle:}
By Jensen's Inequality, the sum of a convex function is minimized when all input variables are perfectly equal:
\begin{equation}
    \mathcal{A}_{\text{side}} = \sum_{k=1}^3 \text{arctanh}\left(\frac{J_{\text{leak}}}{\tau_k}\right) \ge 3 \text{arctanh}\left(\frac{J_{\text{leak}}}{\bar{\tau}}\right).
\end{equation}
Substituting this minimum affinity into the flux-coupling formula Eq.~\eqref{eq:SI_Inhib_derivation_step}, we obtain the minimum for the production flux:
\begin{equation}
    J_{\text{cat}}^{\text{min}} = J_{\text{leak}} + \tau_{01} \tanh\left( 3 \text{arctanh} \frac{J_{\text{leak}}}{\bar{\tau}} \right).
\end{equation}
This confirms that the inhibitor achieves its maximal efficacy (minimal $J_{\text{cat}}$) when the kinetic traffic is balanced across the entire inhibitor cycle.\\

\textit{(B) The Upper Bound ($J_{\text{cat}}^{\text{max}}$) --- The Bottleneck Limit:}
Conversely, the maximum of a convex function under a sum constraint occurs at the boundaries of the feasible domain. For $J_{\text{cat}}$ to be maximized, the sideway traffic distribution must be maximally uneven.\\

Consider the limit where one transition in the inhibitor cycle, say $\tau_1$, becomes a kinetic bottleneck such that $\tau_1 \to J_{\text{leak}}^+$. In this limit:
\begin{enumerate}
    \item The local resistance of that edge diverges: $\text{arctanh}(J_{\text{leak}}/\tau_1) \to \infty$.
    \item Consequently, the total sideway resistance diverges: $\mathcal{A}_{\text{side}} \to \infty$.
    \item Using the asymptotic property of the hyperbolic tangent, $\lim_{x \to \infty} \tanh(x) = 1$.
\end{enumerate}
Substituting this limit into Eq.~\eqref{eq:SI_Inhib_derivation_step}, we derive the upper bound for production:
\begin{equation}
    J_{\text{cat}}^{\text{max}} = J_{\text{leak}} + \tau_{01}.
\end{equation}
This upper bound reflects the state of the inhibitor loop: a bottleneck severs the compensatory leak, freeing the production flux to its maximum capacity dictated by the primary binding traffic $\tau_{01}$.\\

We numerically validate these lower and upper bounds of $J_{\text{cat}}$ in the main text by log-uniformly sampling $10^5$ networks with transition rates spanning six orders of magnitude (from $10^{-3}$ to $10^3$) under the futile loop constraint.\\

\end{document}